\documentclass[%
reprint,
superscriptaddress,
frontmatterverbose,
amsmath,amssymb,
prb,
]{revtex4-1}

\usepackage{amsfonts,amsmath,amssymb,amsthm}

\usepackage{graphicx}% Include figure files
\usepackage{dcolumn}% Align table columns on decimal point
\usepackage{bm}% bold math
\usepackage{hyperref}

\hypersetup{
	pdfnewwindow=true,      % links in new window
	colorlinks=true,       % false: boxed links; true: colored links
	linkcolor=blue,          % color of internal links (change box color with linkbordercolor)
	citecolor=blue,        % color of links to bibliography
	filecolor=blue,      % color of file links
	urlcolor=blue,          % color of external links
}

\usepackage{color,xcolor, ulem}
\usepackage{amsmath}

\usepackage{lineno}
\makeatother

\newcommand{\be}{\begin{equation}}
\newcommand{\ee}{\end{equation}}

\usepackage{braket}
\usepackage{array}
\usepackage{booktabs}
\usepackage{multirow} 
\newif\ifdraft
\drafttrue
\draftfalse

\begin{document}

% \preprint{APS/123-QED}

\title{A computational method to estimate spin-orbital interaction strength in solid state systems}

\author{Qiangqiang Gu}
\email{guqq@pku.edu.cn}
%\affiliation{International Center for Quantum Materials, School of Physics, Peking University, Beijing 100871, China}
\affiliation{School of Mathematical Science, Peking University, Beijing 100871,China}
\affiliation{AI for Science Institute, Beijing, China}

\author{Shishir Kumar Pandey}
\email{shishir.kr.pandey@gmail.com}
\affiliation{AI for Science Institute, Beijing, China}

\date{\today}% It is always \today, today,

\begin{abstract}
Spin-orbit coupling (SOC) drives interesting and non-trivial phenomena in solid state physics, ranging from topological to magnetic to transport properties. 
Thorough study of such phenomena often require effective models where SOC term is explicitly included. However, estimation of SOC strength for such models mostly depend on the spectroscopy experiments which can only provide a rough estimate. In this work, we provide a simple yet effective computational approach to estimate the on-site SOC strength using a combination of the \textit{ab initio} and tight-binding calculations. We demonstrate the wider applicability and high sensitivity of our method considering materials with varying SOC strengths and the number of SOC 
active ions. 
The estimated SOC strengths agree well with the proposed values in literature lending support to our methodology. 
This simplistic approach can readily be applied to a wide range of materials. 

\end{abstract}

%Use showkeys class option if keyword

\maketitle

\section{Introduction}
The spin-orbit coupling (SOC), a relativistic interaction couples electronic spin (\bm{$S$}) and its 
orbital momentum (\bm{$L$}) in an atom, causes the splitting of electronic terms into multiplets
with total angular momentum \bm{$J$} = \bm{$L$} + \bm{$S$}. 
Dependence of SOC strength $\lambda$ on atomic number $Z$ as $\sim$ $Z^4$  make this interaction non-perturbative and crucial 
in heavier elements for accurate description of their electronic structure. 
When these elements put in a crystal, the situation becomes complex as other splitting terms like crystal field (CF) takes part in determining the electronic structure of a solid. 
Nevertheless, SOC still manifests its vital role in materials in the form of various interesting physical phenomena.
Some of the examples are magnetic anisotropy~\cite{maganis2008, maganis2018, maganis2015}, spin current~\cite{rmp_soc}, anomalous Hall effect~\cite{ahe2013, apl2018, prl2015, apl2013} and the topological properties of materials originating from SOC effects~\cite{koning2007, hsieh2008, hsieh2009, hasan2009, chen2009,Wan2011,wengWeyl2015,soluyanov2015}. 
Most recently, in strongly correlated materials, interplay of SOC with other electronic interactions like Coulomb interaction and crystal fields have given rise to exotic phases like unconventional superconductivity~\cite{sc1, sc2, sc3, sc4} and  Kitaev interactions for realization of quantum spin liquid state~\cite{qsl1, qsl2}.
Additionally, in 2D materials such as transition metal dichalcogenides (TMDCs),
SOC effects can be crucial for possible valleytronic and spintronic applications~\cite{tmd}.
Given that SOC can strongly modify the electronic structure near the Fermi level resulting in drastic changes in charge carrier mobility and transport properties, its implications in the context of device based applications are also numerous. 
Thus, SOC effects are imperative for electronic structure modeling of materials in condensed matter physics.

From computation perspective of materials modeling, Density Functional Theory (DFT) based \textit{ab initio} methods have gained wide acceptance for their high transferability and relatively high accuracy. SOC effects can be self-consistently included within this approach.
However, in many cases such as  strong-correlated materials or large-scale systems, application of DFT becomes infeasible.
A work around in such cases is construction of an
effective model e.g. tight-binding (TB), multiband Hubbard-Kanamori\cite{hubbkan}
or Kane-Mele models~\cite{kanem},
which can closely describe materials properties of interest.
Such models have further advantages like their low computation cost,
flexibility to explicitly include additional interactions and vary their strengths to examine the 
corresponding effects on materials properties.  
As a specific example, one can consider the Hamiltonian $H$ = $H_\text{TB}$ + $H_\text{CF}$ + $H_\text{soc}$ + $H_\text{int}$ as an example, 
where first, second and third terms correspond to CF, SOC and Hubbard-Kanamori interaction terms. 
This Hamiltonian has been used to estimate the magnetic interactions in so called spin-orbit coupling assisted Mott insulators~\cite{srrho, coprb}.
One can then vary strength of interactions like $\lambda$ in the SOC term 
($H_{\text{soc}}$ = $\sum_i \lambda_i \bm L_i \cdot \bm S_i$, $i$ is atomic index) 
or intra/inter-orbital Hubbard term ($U$) and/or 
Hund's interaction ($J_\text{H}$) within $H_\text{int}$ term~\cite{coprb}. Techniques such as constrained random-phase–approximation (cRPA) are 
available to provide estimates of $U$ and $J_\text{H}$~\cite{crpa1,crpa2,crpa3}, while hopping ($H_\text{TB}$) and CF terms ($H_\text{CF}$) can been estimated from $ab$ $initio$ calculation with Wannierization procedure~\cite{Marzari2012,coprb, srrho}. 

However, for estimation of $\lambda$, one mostly relies on the experiments like low temperature transport spectroscopy, absorption spectroscopy, electroreflectance measurements. 
Such an estimation of $\lambda$, in many cases, may be erroneous. 
It is because, firstly, spectral features of a solid used to estimate $\lambda$ may not be solely emerging from SOC effects and more often than not influenced by other structural aspects. 
This can lead to an ``effective'' SOC in materials. This is consistent with the fact that in solids with moderate SOC interaction, the atomistic approach to determine strength of SOC cannot be straightforwardly applied. 
For example, as shown by one of us~\cite{coprb} that experimental estimation from SOC assisted excitations brings 
$\lambda$ = 15 meV for Co$^{+2}$ ions in cobaltates and yet our computationally estimated value $\lambda$ = 65 meV reproduces experimental features quite well. 
Secondly, these structural features (e.g. trigonal or tetragonal distortion of an octahedra) may have large variation in different materials. This imply that effective SOC may significantly
vary in different materials. 
However, in the absence of optical measurements for a particular material, one often has to choose value of $\lambda$ from a broad range available from experiments on other materials belonging 
to similar class rather than material of interest itself. 
Given that $\lambda$ depends on the ionic and spin state of corresponding atom in a solid, such an approach might be amiss or erroneous.
Several attempts have been made in the past including some very recent ones to validate and 
estimate the on-SOC interaction in isolated atoms, monoatomic crystals and 
binary compounds~\cite{soc1,soc2,soc3,soc4}. However, a general procedure for estimation of
$\lambda$ which can readily be applied to any material of interest is still missing.  
Computationally, \textit{ab initio} calculations with Wannierization procedure~\cite{soc3, Marzari2012} can produce the 
models that contains the SOC effect. However, in this case, the obtained Wannier orbitals are not necessarily the eigenstates of $\hat{S}_z$ and so, it will be cumbersome to extract $\lambda$ from such models making their analysis and transferability an issue.

In this paper, we present a simple yet effective method to obtain the strength of SOC 
for any number of atomic species present in a material. Our method is based on a combination of $ab$ $initio$ and Wannier function based tight-binding models (dubbed TBSOC~\cite{qq} hereafter). 
In this method, we construct a full SOC Hamiltonian defined in the $|\ell_z, s_z\rangle$ basis, which is then used to fit the $ab$ $initio$ band structure where the SOC effect is self-consistently introduced. SOC strength is then estimated by fitting the $ab$ $initio$ eigenvalues within our model. 
After introducing the methodology, TBSOC is then tested considering various materials ranging from having weak (3$d$) to strong (5$d$) single SOC active ions. We also consider an example of a topological material with two SOC active ions. 
We show that TBSOC works well on these broad range compounds and the estimated SOC strength is in close agreement with the values found in literature. We also show that the choice of local coordinate system does not affect the result which can further be utilized for additional analysis.
These results demonstrate that our TBSOC method offers a direct and convenient way to estimate 
the strength of SOC in materials.

%This method is proved to work well for all the materials ranges from  weak (3d) to strong (5d) SOC materials as well as the topological materials. In the meantime, it works well 
 
 %supports to obtained the SOC interaction strength defined in the $|s_z\rangle$ basis for the non-SOC Wannier tight-binding (TB) models. After introducing the algorithm , TBSOC is tested using various materials, including Iridates (5d), RuCl3(4d), Cobaltates (3d) and TaAs (topological).
 
%\sout{In these models, the SOC strength need to be explicitly treated, especially for the strong correlated materials where the mean-field techniques fails. For example, add some example where the SOC need to be explicitly treated. In these cases,  one need to construct the effective and accurate models, which usually includes the parameters of hopping, crystal fields, Hubbard and SOC interactions. The hoppings and crystal fields can be directly calculated from the \textit{ab initio} method with Wannierization procedure. The Hubbard term can be obtained from crpa calculations or fitting with experiment data. As for the SOC strength, traditionally, one can obtain it from the optical spectral experiments. However, limited by the accuracy of the experimental observation, the obtained SOC strengths are mostly a general range instead  of precise values. } 
%In addition, not all the elements have been experimentally detected. For these elements, the SOC strengths are taken as the values from the Periods in the periodic table

\section{Methodology}\label{methods}

In this section, we first discuss the theoretical background required for implementation of SOC strength followed by methodological procedure discussed next. 

\subsection{Theoretical background}\label{theory}

The TB Hamiltonian for materials can be described in the form as,
\begin{equation}
	H_\text{TB}= \sum_{i\alpha,j\beta} T_{i\alpha,j\beta}c^{\dagger}_{i\alpha}c_{j\beta} + c.c.
\label{tb}
\end{equation} 
where the $T_{i\alpha,j\beta}$ are the hopping elements (when $i\neq j$) or onsite element (when $i = j$) between the basis orbitals $|i\alpha\rangle$ and $|j\beta\rangle$. Here $i$, $j$ are the atomic indices and $\alpha$, $\beta$ label the orbitals basis functions.
These basis functions can be described in terms of either complex spherical harmonic 
($Y_\ell^m$) or real/cubic harmonic functions ($X_{\ell m}$) and the two bases are 
interchangeable using the following unitary transformations. 

\begin{equation}
X_{\ell m} = 
\begin{cases}
\frac{i}{\sqrt{2}}\left(Y_{\ell}^{-|m|}-(-1)^m Y_{\ell}^{|m|}\right) & m<0 \\ 
Y_{\ell}^0 & m=0 \\ 
\frac{1}{\sqrt{2}}\left(Y_{\ell}^{-|m|}+(-1)^m Y_{\ell}^{|m|}\right) & m>0
\end{cases}
\label{trans}
\end{equation}

Above, $\ell$ is the angular momentum or azimuthal quantum number and $m$ is the magnetic quantum number.
However, for efficient computational implementation, the real/cubic harmonic form of orbital basis functions are often considered, functional forms of which are listed in Table.~\ref{tab:realhmnics}. 

\begin{table}[!ht]
 \centering
\caption{The real/cubic harmonics basis functions $X_{\ell m}$.}
\begin{tabular}{ccl}
\hline \hline 
%\specialrule{0em}{0pt}{2pt}
$\ell$ &  &  ~~~~~~~~~~~~~~~~~~~$X_{\ell m}$ \\
%\specialrule{0em}{0pt}{2pt}
\hline
\specialrule{0em}{0pt}{3pt}
$\ell=0$ & & $X_{0,0} ~~= |s\rangle~~~~~~~=\sqrt{1/4\pi}$\\
\specialrule{0em}{0pt}{3pt}

\hline
\specialrule{0em}{0pt}{3pt}

\multirow{3}*{$\ell=1$} & &  $X_{1, -1} = \left|p_y\right\rangle~~~~~=\sqrt{3 / 4 \pi}\cdot y/r$\\
\specialrule{0em}{0pt}{1pt}
~ & & $X_{1, 0} ~~= \left|p_z\right\rangle ~~~~~=\sqrt{3 / 4 \pi}\cdot z/r$\\
\specialrule{0em}{0pt}{1pt}
~ & & $X_{1, 1} ~~= \left|p_x\right\rangle~~~~~=\sqrt{3 / 4 \pi}\cdot x/r$\\
\specialrule{0em}{0pt}{3pt}

\hline
\specialrule{0em}{0pt}{3pt}
\multirow{5}*{$\ell=2$} & & $X_{2, -2} = \left|d_{xy}\right\rangle~~~~=\sqrt{15 / 4 \pi} \cdot xy/r^2 $\\
\specialrule{0em}{0pt}{1pt}
~ & & $X_{2, -1} = \left|d_{yz}\right\rangle~~~~=\sqrt{15 / 4 \pi} \cdot yz/r^2 $\\
\specialrule{0em}{0pt}{1pt}
~ & & $X_{2, 0} ~~= \left|d_{z^2}\right\rangle~~~~=\sqrt{5 /16\pi} \cdot (3z^2-r^2)/r^2 $\\
\specialrule{0em}{0pt}{1pt}
~ & & $X_{2, 1} ~~= \left|d_{xz}\right\rangle~~~~=\sqrt{15 / 4 \pi} \cdot xz/r^2 $\\
\specialrule{0em}{0pt}{1pt}
~ & & $X_{2, 2} ~~= \left|d_{x^2-y^2}\right\rangle=\sqrt{15 /16 \pi} \cdot (x^2-y^2)/r^2 $\\
\specialrule{0em}{0pt}{3pt}
\hline \hline
\end{tabular}
\label{tab:realhmnics}
\end{table}

%$|s \pm\rangle=\sqrt{1 / 4 \pi}|\pm\rangle$ & $\left|p_1 \pm\right\rangle=\sqrt{3 / 4 \pi} f_1(r) x|\pm\rangle$ & $\left|d_1 \pm\right\rangle=\sqrt{5 / 16 \pi} f_2(r) x y|\pm\rangle$ \\
%& $\left|p_2 \pm\right\rangle=\sqrt{3 / 4 \pi} f_1(r) y|\pm\rangle$ & $\left|d_2 \pm\right\rangle=2 \sqrt{15 / 16 \pi} f_2(r) y z|\pm\rangle$ \\
%& $\left|p_3 \pm\right\rangle=\sqrt{3 / 4 \pi} f_1(r) z|\pm\rangle$ & $\left|d_3 \pm\right\rangle=2 \sqrt{15 / 16 \pi} f_2(r) z x|\pm\rangle$\\
%& & $\left|d_4 \pm\right\rangle=\sqrt{15 / 16 \pi} f_2(r)\left(x^2-y^2\right)|\pm\rangle$\\
%& & $\left|d_5 \pm\right\rangle=\sqrt{5 / 16 \pi} f_2(r)\left(3 z^2-r^2\right)|\pm\rangle$\\

The SOC interaction that couples orbital momentum with that of spin can be accurately approximated by a local ``atomic'' contribution 
of form,
\begin{equation}
%H_\text{soc} =\lambda \bm{L}\cdot \bm{S}  =\lambda \left[ L_zS_z + \frac{1}{2}\left(L_+ S_- + L_-S_+\right)\right]
H_\text{soc} = \sum_i \lambda \bm L_i \cdot \bm S_i =\lambda \left[ \hat{L}_z\hat{S}_z + \frac{1}{2}\left(\hat{L}_+ \hat{S}_- + \hat{L}_-\hat{S}_+\right)\right]
%H_\text{soc}= \sum_i \lambda_i \bm{l_i} \cdot \bm{s_i}
\label{rl}
\end{equation} 
$\bm L_i$ and $\bm S_i$ are the angular and spin momentum operator, and $\hat{L}_\pm$, $\hat{S}_\pm$ are the corresponding raising and lowering operators. Since we are discussing the on-site opertaors 
for each site in the crystal, index $i$ is drooped in the right hand side of Eq.~\ref{rl}. 
The operation of orbital angular momentum operators on the complex spherical harmonics $Y_\ell^m$ yields,  
\begin{equation}
\begin{gathered}
L_{\pm} Y_\ell^m=\hbar \sqrt{\ell(\ell+1)-m(m \pm 1)} Y_\ell^{m \pm 1} \\
L_z Y_\ell^m =\hbar m Y_\ell^m .
\end{gathered}
\label{operation}
\end{equation}
The same is also applicable for spin operators.

From Eqs.~\ref{trans}--\ref{operation}, one can obtain the SOC Hamiltonian matrix in the Hilbert subspaces for $p$ and $d$ orbitals
considering the basis functions $\alpha$/$\beta$  = [$p_z,p_x,p_y$]  and 
[$d_{z^2},d_{xz},d_{yz},d_{x^2-y^2}, d_{xy}$] respectively. 
One should note that we haven't considered spin index in the TB Hamiltonian given in Eq.~\ref{tb} and a full Hamiltonian 
$\mathcal{H} = H_\text{TB} + H_\text{soc}$ can be constructed as $\mathcal{H}= \mathcal{I}_2 \otimes H_\text{TB} + H_{\text{soc}}$, 
where $\mathcal{I}_2$ is the two-dimensional identity matrix and $\otimes$ is the Kronecker product. 
The complete
basis then becomes, [$p_{z\uparrow}$, $p_{x\uparrow}$, $p_{y\uparrow}$, $p_{z\downarrow}$,  
 $p_{x\downarrow}$, $p_{y\downarrow}$]  and 
[$d_{z^2\uparrow}$, $d_{xz\uparrow}$, $d_{yz\uparrow}$, $d_{x^2-y^2\uparrow}$, $d_{xy\uparrow}$, 
$d_{z^2\downarrow}$, $d_{xz\downarrow}$, $d_{yz\downarrow}$, $d_{x^2-y^2\downarrow}$, $d_{xy\downarrow}$]. 
Thusly constructed  $H_\text{soc}$ for $p$ and $d$ orbitals are given in Eq.~\ref{eq:Msoc_p} and Eq.~\ref{eq:Msoc_d}. Here, 
$\lambda_p$ and $\lambda_d$ are the SOC strength of atomic 
$p$ and $d$ orbitals.  

\begin{widetext}
\begin{equation}
 H^{p}_{\text{soc}}=
 \frac{\lambda_{p}}{2}
 \left(
 \begin{array}{cccccc}
 0 &  0 &   0 &   0 &  -1 &  i \\ 
 0 &  0 &  -i &   1 &   0 &  0 \\  
 0 &  i &   0 &  -i &   0 &  0 \\  
 0 &  1 &   i &   0 &   0 &  0 \\ 
-1 &  0 &   0 &   0 &   0 &  i \\ 
-i &  0 &   0 &   0 &  -i &  0 
\end{array}
\right)
\label{eq:Msoc_p}
\end{equation}

%\begin{small}
\begin{equation}
H^{d}_{\text{soc}}=
\frac{\lambda_{d}}{2}
\left(
\begin{array}{cccccccccc}
         0 & 0 & 0 & 0 & 0 & 0 & -\sqrt3 & i\sqrt3 & 0 & 0 \\  
         0 & 0 & -i & 0 & 0 & \sqrt3 & 0 & 0 & -1 & i \\  
         0 & i & 0 & 0 & 0 & -i\sqrt3 & 0 & 0 & -i & -1 \\  
         0 & 0 & 0 & 0 & -2i & 0 & 1 & i & 0 & 0 \\  
         0 & 0 & 0 & 2i & 0 & 0 & -i & 1 & 0 & 0 \\  
         0 & \sqrt3 & i\sqrt3 & 0 & 0 & 0 & 0 & 0 & 0 & 0 \\  
   -\sqrt3 & 0 & 0 & 1 & i & 0 & 0 & i & 0 & 0 \\  
  -i\sqrt3 & 0 & 0 & -i & 1 & 0 & -i & 0 & 0 & 0 \\  
         0 & -1 & i & 0 & 0 & 0 & 0 & 0 & 0 & 2i \\  
         0 & -i & -1 & 0 & 0 & 0 & 0 & 0 & -2i & 0   
\end{array}
\right)
\label{eq:Msoc_d}
\end{equation}
%\end{small}
\end{widetext}

%$(|p_z \uparrow\rangle, |p_x \uparrow\rangle,|p_y \uparrow\rangle,|p_z \downarrow\rangle,|p_x \downarrow\rangle,|p_y \downarrow\rangle)$ and $H^{d}_{\text{soc}}$ is defined basis on $($d_z^2,d_{xz},d_{yz},d_{x^2-y^2}, d_{xy}$)$
%(|\uparrow\rangle,|\downarrow\rangle)\bigotimes 

\subsection{Implementation}\label{impl}

Having obtained the $H_\text{SOC}$ in previously, as the next step, we now describe the complete process for the extraction 
of SOC strength in solid state materials. The procedure can be divided into mainly three sequential $steps$ given below. 

\begin{enumerate}
    \item Non-spin polarized $ab$ $initio$ band structure calculation and its Wannier based TB Hamiltonian ($H_\text{TB}$).
    \item Self-consistently SOC included $ab$ $initio$ band structure calculation.
    \item Obtain the optimized SOC strength $\lambda$ by fitting the band structure of $step$ 2 with Hamiltonian $H_\text{TB}$ 
     + $H_\text{SOC}$ (of Section~\ref{theory}). 
\end{enumerate}
In $step$ 3, the derivatives free  Nelder-Mead optimization algorithm ~\cite{NelderMead1965,Gao2012} is used for error minimization when fitting 
the $ab$ $initio$ eigenvalues to obtain the SOC strength $\lambda$.

\subsection{$Ab$ $initio$ calculations} 
In  $steps$ 1 and 2 in the previous section, the $ab$ $initio$ band structure
calculations can be performed with any of the available DFT packages with an interface to 
Wannier90~\cite{wannier90}. 
In this paper, we choose Vienna $ab$ $initio$ simulation package~\cite{Kresse} for the calculations.
Projector-augmented wave method~\cite{paw1, paw2}
implemented within VASP (version-5.4.4) with the generalized-gradient
approximation (GGA) within Perdew-Burke-Ernzerhof
framework~\cite{PBE} is used and the energy convergence criteria in our self-consistent calculations is when energy difference between successive steps was better than 10$^{-5}$/unit cell. Details of the 
plane wave energy cutoff and $k$-grid used for specific materials is provided in their
respective sections.

To summarise our proposed methodology, our TBSOC program takes in a Wannier function based 
TB model ($H_\text{TB}$) and constructs the full Hamiltonian $\mathcal{H} = H_\text{TB} + H_\text{soc}$ by adding the on-site $H_\text{soc}$ term. 
Starting with a random guess initial value of SOC strength $\lambda$, we fit 
the eigenvalues of SOC included $ab$ $initio$ band structure with $\mathcal{H}$. Error minimization 
of the fit then  leads to corresponding $\lambda$ for the atomic species of interest in a material. 

\section{Applications}
Having described the methodology in detail, in this section we show 
the application of TBSOC on some transition metal compounds as well as 
on a topological material. The transition metal compounds considered here are typical examples for magnetic materials 
belonging to 3$d$, 4$d$ and 5$d$ class, whose SOC strength 
 ranges from weak to strong. The topological material (TaAs) we consider here is a typical example of non-magnetic and inversion symmetry breaking Weyl semi-metal. The interesting magnetic or topological properties originating from SOC in these materials make them suitable candidates to test our method.

\subsection{The case of transition metal compounds}

In this section, as case studies, we demonstrate the estimation of
$\lambda$ in the various transition metal compounds. The choice of materials 
here is based on three points. 
First, keeping in mind that SOC strength 
$\lambda$  depends on the number of valence electrons (the ionic state of the atom species of interest in a compound). Hence, we choose  
Na$_2$TeCo$_2$O$_6$ (Co$^{+2}$-3$d^7$) and 
$\alpha$-RuCl$_3$ (Ru$^{+3}$-4$d^5$) as our example systems. 
Second, given a fixed number of valence electrons, $\lambda$ changes with 
atomic number of atomic species of interest. 
Hence, we show a comparison between iso-electronic $\alpha$-RuCl$_3$
(Ru$^{+3}$-4$d^5$) and Na$_2$IrO$_3$ (Ir$^{+4}$-5$d^5$). 
Third, our choice is also based on the drastic variation of the SOC strength 
in these materials, which in different cases compete with other interactions 
likes the trigonal/tetragonal crystal field splitting. 
Magnitude of such splitting in these materials is believed to be in the 
range 10 -- 40 meV~\cite{winterprb} and the SOC strength 
$\lambda$ is either comparable or larger than the range of these splitting in 
the aforementioned compounds. 
This show the effectiveness of our methodology apart from its diverse applicability.

\begin{figure}[ht!]
\centering
\includegraphics[width=7.5 cm]{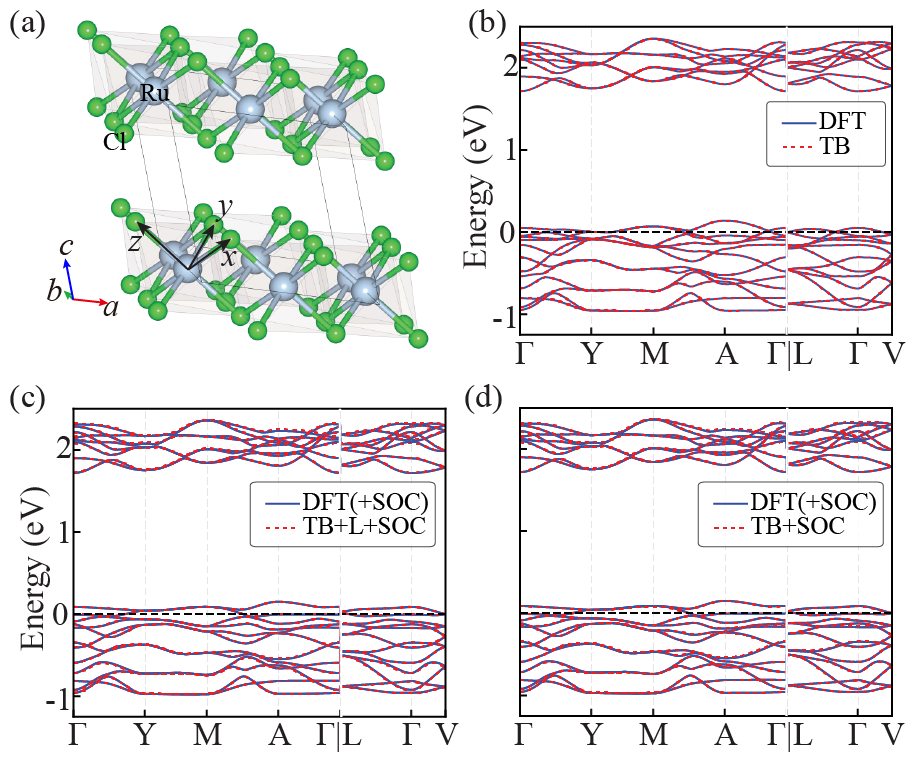}
\caption{(a) Crystal structure of $\alpha$-RuCl$_3$. 
Light blue and green balls represent Ru and Cl atoms respectively. $a$, 
$b$ and $c$ are the crystallographic axes while $x$, $y$ and $z$ are the 
local octahedral coordinate axes. Edge-shared Ru-Cl$_6$ octahedron are evident. 
(b) Wannier interpolation of the non-spin polarized 
$ab$ $initio$ band structure of $\alpha$-RuCl$_3$ to obtain $H_\text{TB}$. 
All the five $d$ orbitals of Ru are considered in the basis.
 (c) and (d) show the fitting of SOC included $ab$ $initio$ band structure 
with $H_\text{TB}$ + $H_\text{SOC}$ in two cases, with and with 
local octahedral coordinate system respectively. Fermi energy is set to zero. } 
\label{fig:rcl}
\end{figure}

As the first example, we showcase the results for $\alpha$-RuCl$_3$. 
For the $ab$ $initio$ calculations, we have considered the experimentally 
observed crystal structure with monoclinic space group $C$/$2m$~\cite{rcl_exp1,rcl_exp2,rcl_exp3}, crystal 
structure of which is shown in Fig.~\ref{fig:rcl} (a). 
We used plane wave energy cutoff 500 eV and $\Gamma$-center $k$-grid of 6$\times$3$\times$6 
in our DFT self-consistent calculations.
The layered structure with edge-shared Ru-Cl$_6$ octahedron
are evident from this figure. As described in Section~\ref{methods}, in the first
step, we construct a TB Hamiltonian ($H_\text{TB}$) using Wannier 
interpolation of the non-spin polarised $ab$ $initio$ band structure and the 
plot is shown in Fig.~\ref{fig:rcl}(b). One  can see a very good agreement 
between the two band structures from this figure.
This gives us confidence to proceed to the second step in which 
we calculate the $ab$ $initio$ band structure with SOC included 
at the self-consistent level. We then fit this relativistic band structure 
with $H_\text{TB}$ after adding the on-site SOC $H_\text{TB}$ term. 
A plot of this fitting procedure is shown in Fig.~\ref{fig:rcl}(c). 
We have emphasize here that in both the cases in
Fig.~\ref{fig:rcl}(b) and (c) we use local octahedral coordinate
system shown in Fig.~\ref{fig:rcl}(a) to obtain $H_\text{TB}$. 
It is also possible to obtain the estimate of $\lambda$ is the global 
crystallographic coordinate axes and fitting procedure is not affected by 
the choice of coordinate settings. This is evident from Fig.~\ref{fig:rcl}(d)
where the $H_\text{TB}$ obtained in global coordinates is used for the fitting.
The estimated value of $\lambda$ in both the cases is $\sim$ 0.120 eV. 
This value of $\lambda$, though in close agreement, is slightly larger by 20 meV from previously experimentally estimated value~\cite{rcl_soc_exp1,rcl_soc_exp2}. As explained in the introduction, 
underestimation of $\lambda$ from experiments may come from other structural distortions 
at play.
We emphasis here that further analysis of tight binding model, often required to understand the electronic properties of the
material, is not straight forward when global
coordinate is used in the honeycomb lattice systems. Hence, we recommend the use of local coordinate system in such 
cases.

\begin{figure}[ht]
\centering
\includegraphics[width=7.5 cm]{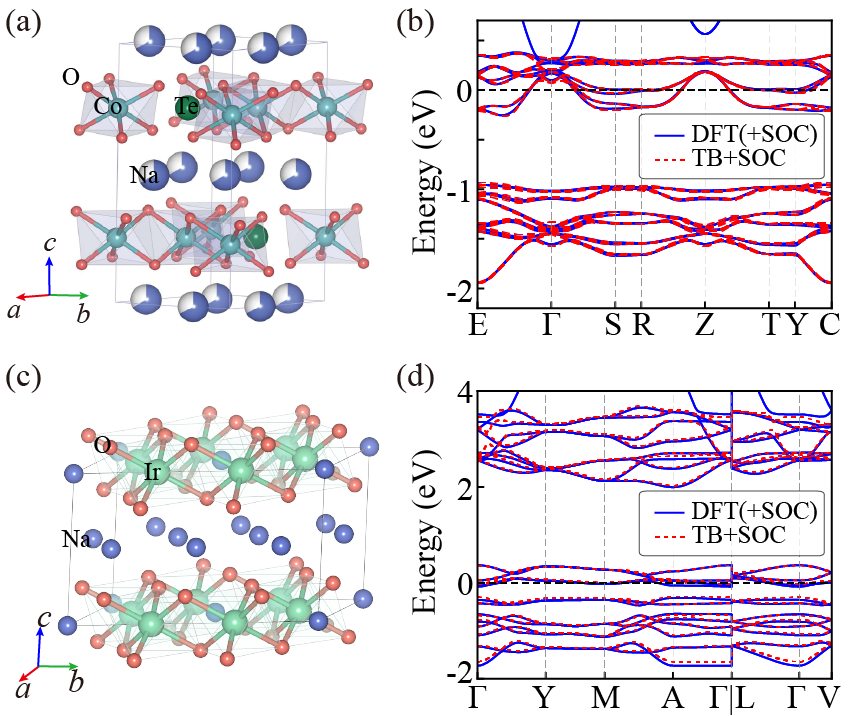}
\caption{(a) and (c) Crystal structure of Na$_3$TeCo$_2$O$_6$ and Na$_2$IrO$_3$. 
The white-blue ball in (a) represents partial occupancy of the Na sites. $a$, 
$b$ and $c$ are the crystallographic axes. Cyan/green and red balls represent 
Co/Ir and oxygen atoms respectively. Edge-shared Co/Ir-O$_6$ octahedron 
are evident in (a) 
and (c). (b) and (d) show the fitting of SOC included $ab$ $initio$ band structure
with a tight binding model after including on-site spin-orbit coupling term for 
Na$_3$TeCo$_2$O$_6$ and Na$_2$IrO$_3$ respectively. All the five $d$ orbitals are considered in tight binding basis in both these cases. Fermi energy is set to zero. } 
\label{fig:3d5d}
\end{figure}

In order to demonstrate the sensitivity of our method, following the same 
procedure, we estimate $\lambda$ 
for two more cases, namely for Na$_2$TeCo$_2$O$_6$ and Na$_2$IrO$_3$. 
For these two cases, strength of SOC varies drastically for Co$^{+2}$ and Ir$^{+4}$ ions. 
We consider the crystal structure of Na$_2$TeCo$_2$O$_6$ with 
space group $P6_322$~\cite{co_exp1,co_exp2,co_exp3,co_exp4,co_exp5,co_exp6}
and Na$_2$IrO$_3$ with $C2/m$~\cite{ir_exp1,ir_exp2,ir_exp3} which are shown in 
Fig.~\ref{fig:3d5d}(a) and (c) respectively. 
We used plane wave energy cutoff
550 eV and $\Gamma$-center $k$-grid of 8$\times$8$\times$4 for Na$_2$TeCo$_2$O$_6$ and 
 8$\times$6$\times$8 for Na$_2$IrO$_3$ in our DFT self-consistent calculations. 
Fitting of the SOC included 
$ab$ $initio$ band structure in these two materials is shown in
Fig.~\ref{fig:3d5d}(b) and (d). One can see a very good fitting is obtained in 
both the cases. Estimated values of $\lambda$ are 0.065 eV for 
Na$_2$TeCo$_2$O$_6$ and 0.380 eV for Na$_2$IrO$_3$ which matches well with the previous experimental 
estimation~\cite{ir_soc_exp}.

In this section, 
we have considered examples in which there were only one spin-orbit coupling 
activated transition metal ion present. However, there can be more than 
one atomic species in a material with active spin-orbit coupling interaction. 
To demonstrated applicability of our method in such cases as well, we consider an example of a topological material, namely TaAs in the next section.

\subsection{The case of topological material}

\begin{figure}[ht!]
\centering
\includegraphics[width=7.5 cm]{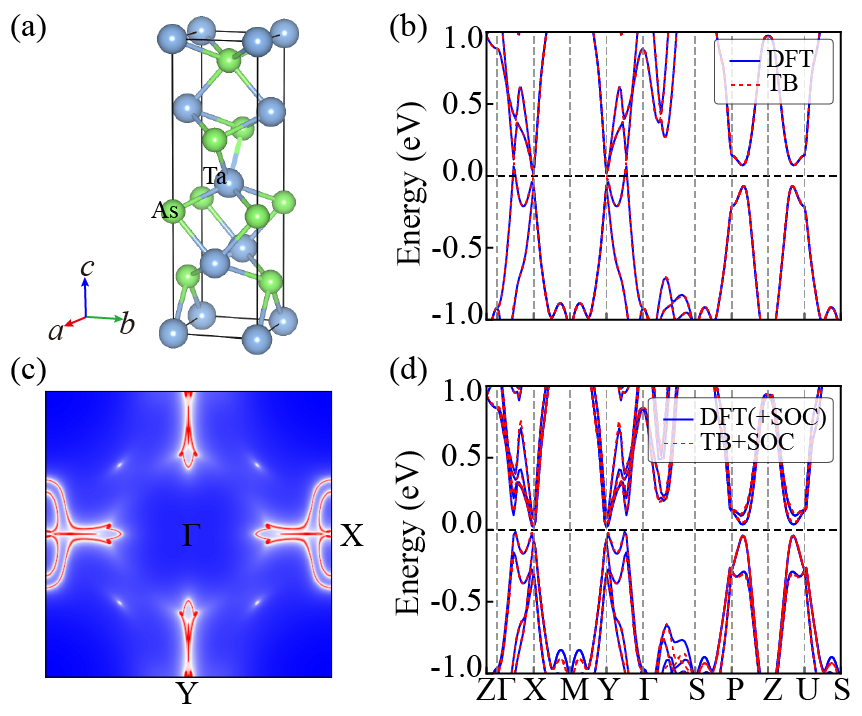}
\caption{(a) The body-centered-tetragonal crystals structure of TaAs. Light blue and green balls represent Ta and As atoms respectively. $a$, $b$ and $c$ are the crystallographic axes.
(b) Wannier interpolation of the non-spin polarized 
$ab$ $initio$ band structure of TaAs to obtain $H_\text{TB}$. 
All the five Ta-$d$ and three As-$p$ orbitals are considered in the basis. (c) The (001) surface states calculated using the $\mathcal{H}$ Hamiltonian. (d) Fitting of the SOC included $ab$ $initio$ band structure with $\mathcal{H}$. Fermi energy is set to zero.}  
\label{fig:taas}
\end{figure}

 TaAs have been theoretically predicted~\cite{wengWeyl2015} and then experimentally proved being the topological Weyl semi-metal~\cite{lvExperimental2015}.
 Its topological properties such as the surface Fermi arcs~\cite{wengWeyl2015}, edge states~\cite{zhengObservation2022}, surface-bulk connectivity ~\cite{inoueQuasiparticle2016}, chiral anomaly\cite{huangchiral2015} and non-linear optical responses\cite{osterhoudt2019}, etc. have been thoroughly studied and are driven by the SOC effects. 
In this material, both, the Ta-$d$ and As-$p$ orbitals are considered to be SOC active atoms.   It crystallizes in a  body-centered-tetragonal structure as shown in Fig.~\ref{fig:taas}(a) with the non-centrosymmetric space group $I4_1md$. To describe this system, one needs to construct the full Hamiltonian $\mathcal{H}$ as explained earlier.  The $H_{\text{TB}}$ model in $\mathcal{H}$ is again obtained from non-spin polarised DFT calculation with the Wannier interpolation procedure. The plane wave energy cutoff is set to be 500 eV and $\Gamma$-center $k$-grid of 10$\times$10$\times$4 are used in the self-consistent DFT calculations.   We considered As-$p$ and Ta-$d$ orbitals as the projection basis functions in the TB model. The non-spin polarized $ab$ $initio$ band structure and its Wannier interpolation are shown are shown in Fig.~\ref{fig:taas}(b). 
One can see that the $H_{\text{TB}}$ in this case also reproduces well the $ab$ $initio$ 
band structure. 
After adding $H_{\text{soc}}$ with parameters $\lambda_p$ and $\lambda_d$ to $H_{\text{TB}}$ forming 
$\mathcal{H}$, we fit the SOC included $ab$ $initio$ band structure with $\mathcal{H}$ in the next 
step. The fit is shown in Fig.~\ref{fig:taas}(d). On can see that the fitting is quite good. 
This brings  $\lambda_p = 0.18$ eV for  As-$p$ and $\lambda_d = 0.21$ eV for Ta-$d$  orbitals, 
which is in the close agreements with the values for Ta atoms as reported in Ref.~\cite{Shanavas2014}.
For lending support to the reliability of our estimated values of $\lambda$'s and thusly obtained full Hamiltonian $\mathcal{H}$, the (001) As terminated surface states of TaAs is calculated using the iterative Green's function ~\cite{sanchoHighly1985}. As shown in Fig.~\ref{fig:taas}(c), a  spoon-shaped surface states around the $X$, and $Y$ points appears on the surface Brillouin zone. Along $\Gamma$-X and $\Gamma$-$Y$ directions, the Fermi-arc states connect and terminates at the Weyl points, which agrees well with the reported results from literature~\cite{wengWeyl2015}. 

\section{Conclusion}
We have presented a computational framework named TBSOC~\cite{qq} to  estimate the strength of SOC in solid state materials. 
Our methodology is based on the combination of \textit{ab initio} and tight-binding methods.  To 
show its wider applicability, as case studies, we have considered transition metal compounds $\alpha$-RuCl$_3$, Na$_2$TeCo$_2$O$_6$ and Na$_2$IrO$_3$ and a topological semi-metal TaAs system. We demonstrate the ability of our TBSOC in well reproducing the self-consistent SOC included 
$ab$ $initio$ band structures. This has been done employing a TB model $H_\text{TB}$ with addition of 
on-site SOC term $H_\text{soc}$. We show that it works well in both the cases, materials with strong and weak SOC interactions, covering  a broad range of materials.
We also show that TBSOC works for both local and global coordinate systems, making its utility smoother  
 for further analysis and employment in successive model calculations. This simplistic approach can readily be applied to a wide range of materials.

%\begin{acknowledgments}
%This work is supported by Foundation of XXX.
%the National Natural Science Foundation of China (Grant No. XXX and No. XXX), the Ministry of Science and Technology of China (Grant No. XXX and No. XXX), and by the Strategic Priority Research Program of Chinese Academy of Science (Grant No. XXX). 
 %We gratefully thank Prof. Ji Feng for providing computation resources.
 
%\end{acknowledgments}

\bibliography{soc.bib}

\begin{thebibliography}{66}%
\makeatletter
\providecommand \@ifxundefined [1]{%
 \@ifx{#1\undefined}
}%
\providecommand \@ifnum [1]{%
 \ifnum #1\expandafter \@firstoftwo
 \else \expandafter \@secondoftwo
 \fi
}%
\providecommand \@ifx [1]{%
 \ifx #1\expandafter \@firstoftwo
 \else \expandafter \@secondoftwo
 \fi
}%
\providecommand \natexlab [1]{#1}%
\providecommand \enquote  [1]{``#1''}%
\providecommand \bibnamefont  [1]{#1}%
\providecommand \bibfnamefont [1]{#1}%
\providecommand \citenamefont [1]{#1}%
\providecommand \href@noop [0]{\@secondoftwo}%
\providecommand \href [0]{\begingroup \@sanitize@url \@href}%
\providecommand \@href[1]{\@@startlink{#1}\@@href}%
\providecommand \@@href[1]{\endgroup#1\@@endlink}%
\providecommand \@sanitize@url [0]{\catcode `\\12\catcode `\$12\catcode
  `\&12\catcode `\#12\catcode `\^12\catcode `\_12\catcode `\%12\relax}%
\providecommand \@@startlink[1]{}%
\providecommand \@@endlink[0]{}%
\providecommand \url  [0]{\begingroup\@sanitize@url \@url }%
\providecommand \@url [1]{\endgroup\@href {#1}{\urlprefix }}%
\providecommand \urlprefix  [0]{URL }%
\providecommand \Eprint [0]{\href }%
\providecommand \doibase [0]{http://dx.doi.org/}%
\providecommand \selectlanguage [0]{\@gobble}%
\providecommand \bibinfo  [0]{\@secondoftwo}%
\providecommand \bibfield  [0]{\@secondoftwo}%
\providecommand \translation [1]{[#1]}%
\providecommand \BibitemOpen [0]{}%
\providecommand \bibitemStop [0]{}%
\providecommand \bibitemNoStop [0]{.\EOS\space}%
\providecommand \EOS [0]{\spacefactor3000\relax}%
\providecommand \BibitemShut  [1]{\csname bibitem#1\endcsname}%
\let\auto@bib@innerbib\@empty
%</preamble>
\bibitem [{\citenamefont {Park}\ \emph {et~al.}(2008)\citenamefont {Park},
  \citenamefont {Wunderlich}, \citenamefont {Williams}, \citenamefont {Joo},
  \citenamefont {Jung}, \citenamefont {Shin}, \citenamefont {Olejn\'{\i}k},
  \citenamefont {Shick},\ and\ \citenamefont {Jungwirth}}]{maganis2008}%
  \BibitemOpen
  \bibfield  {author} {\bibinfo {author} {\bibfnamefont {B.~G.}\ \bibnamefont
  {Park}}, \bibinfo {author} {\bibfnamefont {J.}~\bibnamefont {Wunderlich}},
  \bibinfo {author} {\bibfnamefont {D.~A.}\ \bibnamefont {Williams}}, \bibinfo
  {author} {\bibfnamefont {S.~J.}\ \bibnamefont {Joo}}, \bibinfo {author}
  {\bibfnamefont {K.~Y.}\ \bibnamefont {Jung}}, \bibinfo {author}
  {\bibfnamefont {K.~H.}\ \bibnamefont {Shin}}, \bibinfo {author}
  {\bibfnamefont {K.}~\bibnamefont {Olejn\'{\i}k}}, \bibinfo {author}
  {\bibfnamefont {A.~B.}\ \bibnamefont {Shick}}, \ and\ \bibinfo {author}
  {\bibfnamefont {T.}~\bibnamefont {Jungwirth}},\ }\href {\doibase
  10.1103/PhysRevLett.100.087204} {\bibfield  {journal} {\bibinfo  {journal}
  {Phys. Rev. Lett.}\ }\textbf {\bibinfo {volume} {100}},\ \bibinfo {pages}
  {087204} (\bibinfo {year} {2008})}\BibitemShut {NoStop}%
\bibitem [{\citenamefont {Scherer}\ and\ \citenamefont
  {Andersen}(2018)}]{maganis2018}%
  \BibitemOpen
  \bibfield  {author} {\bibinfo {author} {\bibfnamefont {D.~D.}\ \bibnamefont
  {Scherer}}\ and\ \bibinfo {author} {\bibfnamefont {B.~M.}\ \bibnamefont
  {Andersen}},\ }\href@noop {} {\bibfield  {journal} {\bibinfo  {journal}
  {Physical Review Letters}\ }\textbf {\bibinfo {volume} {121}},\ \bibinfo
  {pages} {037205} (\bibinfo {year} {2018})}\BibitemShut {NoStop}%
\bibitem [{\citenamefont {Wa\ss{}er}\ \emph {et~al.}(2015)\citenamefont
  {Wa\ss{}er}, \citenamefont {Schneidewind}, \citenamefont {Sidis},
  \citenamefont {Wurmehl}, \citenamefont {Aswartham}, \citenamefont
  {B\"uchner},\ and\ \citenamefont {Braden}}]{maganis2015}%
  \BibitemOpen
  \bibfield  {author} {\bibinfo {author} {\bibfnamefont {F.}~\bibnamefont
  {Wa\ss{}er}}, \bibinfo {author} {\bibfnamefont {A.}~\bibnamefont
  {Schneidewind}}, \bibinfo {author} {\bibfnamefont {Y.}~\bibnamefont {Sidis}},
  \bibinfo {author} {\bibfnamefont {S.}~\bibnamefont {Wurmehl}}, \bibinfo
  {author} {\bibfnamefont {S.}~\bibnamefont {Aswartham}}, \bibinfo {author}
  {\bibfnamefont {B.}~\bibnamefont {B\"uchner}}, \ and\ \bibinfo {author}
  {\bibfnamefont {M.}~\bibnamefont {Braden}},\ }\href {\doibase
  10.1103/PhysRevB.91.060505} {\bibfield  {journal} {\bibinfo  {journal} {Phys.
  Rev. B}\ }\textbf {\bibinfo {volume} {91}},\ \bibinfo {pages} {060505}
  (\bibinfo {year} {2015})}\BibitemShut {NoStop}%
\bibitem [{\citenamefont {Sinova}\ \emph {et~al.}(2015)\citenamefont {Sinova},
  \citenamefont {Valenzuela}, \citenamefont {Wunderlich}, \citenamefont
  {Back},\ and\ \citenamefont {Jungwirth}}]{rmp_soc}%
  \BibitemOpen
  \bibfield  {author} {\bibinfo {author} {\bibfnamefont {J.}~\bibnamefont
  {Sinova}}, \bibinfo {author} {\bibfnamefont {S.~O.}\ \bibnamefont
  {Valenzuela}}, \bibinfo {author} {\bibfnamefont {J.}~\bibnamefont
  {Wunderlich}}, \bibinfo {author} {\bibfnamefont {C.~H.}\ \bibnamefont
  {Back}}, \ and\ \bibinfo {author} {\bibfnamefont {T.}~\bibnamefont
  {Jungwirth}},\ }\href {\doibase 10.1103/RevModPhys.87.1213} {\bibfield
  {journal} {\bibinfo  {journal} {Rev. Mod. Phys.}\ }\textbf {\bibinfo {volume}
  {87}},\ \bibinfo {pages} {1213} (\bibinfo {year} {2015})}\BibitemShut
  {NoStop}%
\bibitem [{\citenamefont {Vedyayev}\ \emph
  {et~al.}(2013{\natexlab{a}})\citenamefont {Vedyayev}, \citenamefont
  {Ryzhanova}, \citenamefont {Strelkov},\ and\ \citenamefont
  {Dieny}}]{ahe2013}%
  \BibitemOpen
  \bibfield  {author} {\bibinfo {author} {\bibfnamefont {A.}~\bibnamefont
  {Vedyayev}}, \bibinfo {author} {\bibfnamefont {N.}~\bibnamefont {Ryzhanova}},
  \bibinfo {author} {\bibfnamefont {N.}~\bibnamefont {Strelkov}}, \ and\
  \bibinfo {author} {\bibfnamefont {B.}~\bibnamefont {Dieny}},\ }\href
  {\doibase 10.1103/PhysRevLett.110.247204} {\bibfield  {journal} {\bibinfo
  {journal} {Phys. Rev. Lett.}\ }\textbf {\bibinfo {volume} {110}},\ \bibinfo
  {pages} {247204} (\bibinfo {year} {2013}{\natexlab{a}})}\BibitemShut
  {NoStop}%
\bibitem [{\citenamefont {Zhuravlev}\ \emph {et~al.}(2018)\citenamefont
  {Zhuravlev}, \citenamefont {Alexandrov}, \citenamefont {Tao},\ and\
  \citenamefont {Tsymbal}}]{apl2018}%
  \BibitemOpen
  \bibfield  {author} {\bibinfo {author} {\bibfnamefont {M.~Y.}\ \bibnamefont
  {Zhuravlev}}, \bibinfo {author} {\bibfnamefont {A.}~\bibnamefont
  {Alexandrov}}, \bibinfo {author} {\bibfnamefont {L.}~\bibnamefont {Tao}}, \
  and\ \bibinfo {author} {\bibfnamefont {E.~Y.}\ \bibnamefont {Tsymbal}},\
  }\href@noop {} {\bibfield  {journal} {\bibinfo  {journal} {Applied Physics
  Letters}\ }\textbf {\bibinfo {volume} {113}},\ \bibinfo {pages} {172405}
  (\bibinfo {year} {2018})}\BibitemShut {NoStop}%
\bibitem [{\citenamefont {Matos-Abiague}\ and\ \citenamefont
  {Fabian}(2015)}]{prl2015}%
  \BibitemOpen
  \bibfield  {author} {\bibinfo {author} {\bibfnamefont {A.}~\bibnamefont
  {Matos-Abiague}}\ and\ \bibinfo {author} {\bibfnamefont {J.}~\bibnamefont
  {Fabian}},\ }\href {\doibase 10.1103/PhysRevLett.115.056602} {\bibfield
  {journal} {\bibinfo  {journal} {Phys. Rev. Lett.}\ }\textbf {\bibinfo
  {volume} {115}},\ \bibinfo {pages} {056602} (\bibinfo {year}
  {2015})}\BibitemShut {NoStop}%
\bibitem [{\citenamefont {Vedyayev}\ \emph
  {et~al.}(2013{\natexlab{b}})\citenamefont {Vedyayev}, \citenamefont {Titova},
  \citenamefont {Ryzhanova}, \citenamefont {Zhuravlev},\ and\ \citenamefont
  {Tsymbal}}]{apl2013}%
  \BibitemOpen
  \bibfield  {author} {\bibinfo {author} {\bibfnamefont {A.}~\bibnamefont
  {Vedyayev}}, \bibinfo {author} {\bibfnamefont {M.}~\bibnamefont {Titova}},
  \bibinfo {author} {\bibfnamefont {N.}~\bibnamefont {Ryzhanova}}, \bibinfo
  {author} {\bibfnamefont {M.~Y.}\ \bibnamefont {Zhuravlev}}, \ and\ \bibinfo
  {author} {\bibfnamefont {E.~Y.}\ \bibnamefont {Tsymbal}},\ }\href@noop {}
  {\bibfield  {journal} {\bibinfo  {journal} {Applied Physics Letters}\
  }\textbf {\bibinfo {volume} {103}},\ \bibinfo {pages} {032406} (\bibinfo
  {year} {2013}{\natexlab{b}})}\BibitemShut {NoStop}%
\bibitem [{\citenamefont {König}\ \emph {et~al.}(2007)\citenamefont {König},
  \citenamefont {Wiedmann}, \citenamefont {Brüne}, \citenamefont {Roth},
  \citenamefont {Buhmann}, \citenamefont {Molenkamp}, \citenamefont {Qi},\ and\
  \citenamefont {Zhang}}]{koning2007}%
  \BibitemOpen
  \bibfield  {author} {\bibinfo {author} {\bibfnamefont {M.}~\bibnamefont
  {König}}, \bibinfo {author} {\bibfnamefont {S.}~\bibnamefont {Wiedmann}},
  \bibinfo {author} {\bibfnamefont {C.}~\bibnamefont {Brüne}}, \bibinfo
  {author} {\bibfnamefont {A.}~\bibnamefont {Roth}}, \bibinfo {author}
  {\bibfnamefont {H.}~\bibnamefont {Buhmann}}, \bibinfo {author} {\bibfnamefont
  {L.~W.}\ \bibnamefont {Molenkamp}}, \bibinfo {author} {\bibfnamefont {X.-L.}\
  \bibnamefont {Qi}}, \ and\ \bibinfo {author} {\bibfnamefont {S.-C.}\
  \bibnamefont {Zhang}},\ }\href {\doibase 10.1126/science.1148047} {\bibfield
  {journal} {\bibinfo  {journal} {Science}\ }\textbf {\bibinfo {volume}
  {318}},\ \bibinfo {pages} {766} (\bibinfo {year} {2007})}\BibitemShut
  {NoStop}%
\bibitem [{\citenamefont {Hsieh}\ \emph {et~al.}(2008)\citenamefont {Hsieh},
  \citenamefont {Qian}, \citenamefont {Wray}, \citenamefont {Xia},
  \citenamefont {Hor}, \citenamefont {Cava},\ and\ \citenamefont
  {Hasan}}]{hsieh2008}%
  \BibitemOpen
  \bibfield  {author} {\bibinfo {author} {\bibfnamefont {D.}~\bibnamefont
  {Hsieh}}, \bibinfo {author} {\bibfnamefont {D.}~\bibnamefont {Qian}},
  \bibinfo {author} {\bibfnamefont {L.}~\bibnamefont {Wray}}, \bibinfo {author}
  {\bibfnamefont {Y.}~\bibnamefont {Xia}}, \bibinfo {author} {\bibfnamefont
  {Y.~S.}\ \bibnamefont {Hor}}, \bibinfo {author} {\bibfnamefont {R.~J.}\
  \bibnamefont {Cava}}, \ and\ \bibinfo {author} {\bibfnamefont {M.~Z.}\
  \bibnamefont {Hasan}},\ }\href@noop {} {\bibfield  {journal} {\bibinfo
  {journal} {Nature}\ }\textbf {\bibinfo {volume} {452}},\ \bibinfo {pages}
  {970} (\bibinfo {year} {2008})}\BibitemShut {NoStop}%
\bibitem [{\citenamefont {Hsieh}\ \emph
  {et~al.}(2009{\natexlab{a}})\citenamefont {Hsieh}, \citenamefont {Xia},
  \citenamefont {Qian}, \citenamefont {Wray}, \citenamefont {Meier},
  \citenamefont {Osterwalder}, \citenamefont {Patthey}, \citenamefont
  {Checkelsky}, \citenamefont {Ong}, \citenamefont {Fedorov} \emph
  {et~al.}}]{hsieh2009}%
  \BibitemOpen
  \bibfield  {author} {\bibinfo {author} {\bibfnamefont {D.}~\bibnamefont
  {Hsieh}}, \bibinfo {author} {\bibfnamefont {Y.}~\bibnamefont {Xia}}, \bibinfo
  {author} {\bibfnamefont {D.}~\bibnamefont {Qian}}, \bibinfo {author}
  {\bibfnamefont {L.}~\bibnamefont {Wray}}, \bibinfo {author} {\bibfnamefont
  {F.}~\bibnamefont {Meier}}, \bibinfo {author} {\bibfnamefont
  {J.}~\bibnamefont {Osterwalder}}, \bibinfo {author} {\bibfnamefont
  {L.}~\bibnamefont {Patthey}}, \bibinfo {author} {\bibfnamefont {J.~G.}\
  \bibnamefont {Checkelsky}}, \bibinfo {author} {\bibfnamefont
  {N.}~\bibnamefont {Ong}}, \bibinfo {author} {\bibfnamefont {A.~V.}\
  \bibnamefont {Fedorov}},  \emph {et~al.},\ }\href@noop {} {\bibfield
  {journal} {\bibinfo  {journal} {Nature}\ }\textbf {\bibinfo {volume} {460}},\
  \bibinfo {pages} {1101} (\bibinfo {year} {2009}{\natexlab{a}})}\BibitemShut
  {NoStop}%
\bibitem [{\citenamefont {Hsieh}\ \emph
  {et~al.}(2009{\natexlab{b}})\citenamefont {Hsieh}, \citenamefont {Xia},
  \citenamefont {Qian}, \citenamefont {Wray}, \citenamefont {Meier},
  \citenamefont {Dil}, \citenamefont {Osterwalder}, \citenamefont {Patthey},
  \citenamefont {Fedorov}, \citenamefont {Lin}, \citenamefont {Bansil},
  \citenamefont {Grauer}, \citenamefont {Hor}, \citenamefont {Cava},\ and\
  \citenamefont {Hasan}}]{hasan2009}%
  \BibitemOpen
  \bibfield  {author} {\bibinfo {author} {\bibfnamefont {D.}~\bibnamefont
  {Hsieh}}, \bibinfo {author} {\bibfnamefont {Y.}~\bibnamefont {Xia}}, \bibinfo
  {author} {\bibfnamefont {D.}~\bibnamefont {Qian}}, \bibinfo {author}
  {\bibfnamefont {L.}~\bibnamefont {Wray}}, \bibinfo {author} {\bibfnamefont
  {F.}~\bibnamefont {Meier}}, \bibinfo {author} {\bibfnamefont {J.~H.}\
  \bibnamefont {Dil}}, \bibinfo {author} {\bibfnamefont {J.}~\bibnamefont
  {Osterwalder}}, \bibinfo {author} {\bibfnamefont {L.}~\bibnamefont
  {Patthey}}, \bibinfo {author} {\bibfnamefont {A.~V.}\ \bibnamefont
  {Fedorov}}, \bibinfo {author} {\bibfnamefont {H.}~\bibnamefont {Lin}},
  \bibinfo {author} {\bibfnamefont {A.}~\bibnamefont {Bansil}}, \bibinfo
  {author} {\bibfnamefont {D.}~\bibnamefont {Grauer}}, \bibinfo {author}
  {\bibfnamefont {Y.~S.}\ \bibnamefont {Hor}}, \bibinfo {author} {\bibfnamefont
  {R.~J.}\ \bibnamefont {Cava}}, \ and\ \bibinfo {author} {\bibfnamefont
  {M.~Z.}\ \bibnamefont {Hasan}},\ }\href {\doibase
  10.1103/PhysRevLett.103.146401} {\bibfield  {journal} {\bibinfo  {journal}
  {Phys. Rev. Lett.}\ }\textbf {\bibinfo {volume} {103}},\ \bibinfo {pages}
  {146401} (\bibinfo {year} {2009}{\natexlab{b}})}\BibitemShut {NoStop}%
\bibitem [{\citenamefont {Chen}\ \emph {et~al.}(2009)\citenamefont {Chen},
  \citenamefont {Analytis}, \citenamefont {Chu}, \citenamefont {Liu},
  \citenamefont {Mo}, \citenamefont {Qi}, \citenamefont {Zhang}, \citenamefont
  {Lu}, \citenamefont {Dai}, \citenamefont {Fang}, \citenamefont {Zhang},
  \citenamefont {Fisher}, \citenamefont {Hussain},\ and\ \citenamefont
  {Shen}}]{chen2009}%
  \BibitemOpen
  \bibfield  {author} {\bibinfo {author} {\bibfnamefont {Y.~L.}\ \bibnamefont
  {Chen}}, \bibinfo {author} {\bibfnamefont {J.~G.}\ \bibnamefont {Analytis}},
  \bibinfo {author} {\bibfnamefont {J.-H.}\ \bibnamefont {Chu}}, \bibinfo
  {author} {\bibfnamefont {Z.~K.}\ \bibnamefont {Liu}}, \bibinfo {author}
  {\bibfnamefont {S.-K.}\ \bibnamefont {Mo}}, \bibinfo {author} {\bibfnamefont
  {X.~L.}\ \bibnamefont {Qi}}, \bibinfo {author} {\bibfnamefont {H.~J.}\
  \bibnamefont {Zhang}}, \bibinfo {author} {\bibfnamefont {D.~H.}\ \bibnamefont
  {Lu}}, \bibinfo {author} {\bibfnamefont {X.}~\bibnamefont {Dai}}, \bibinfo
  {author} {\bibfnamefont {Z.}~\bibnamefont {Fang}}, \bibinfo {author}
  {\bibfnamefont {S.~C.}\ \bibnamefont {Zhang}}, \bibinfo {author}
  {\bibfnamefont {I.~R.}\ \bibnamefont {Fisher}}, \bibinfo {author}
  {\bibfnamefont {Z.}~\bibnamefont {Hussain}}, \ and\ \bibinfo {author}
  {\bibfnamefont {Z.-X.}\ \bibnamefont {Shen}},\ }\href {\doibase
  10.1126/science.1173034} {\bibfield  {journal} {\bibinfo  {journal}
  {Science}\ }\textbf {\bibinfo {volume} {325}},\ \bibinfo {pages} {178}
  (\bibinfo {year} {2009})}\BibitemShut {NoStop}%
\bibitem [{\citenamefont {Wan}\ \emph {et~al.}(2011)\citenamefont {Wan},
  \citenamefont {Turner}, \citenamefont {Vishwanath},\ and\ \citenamefont
  {Savrasov}}]{Wan2011}%
  \BibitemOpen
  \bibfield  {author} {\bibinfo {author} {\bibfnamefont {X.}~\bibnamefont
  {Wan}}, \bibinfo {author} {\bibfnamefont {A.~M.}\ \bibnamefont {Turner}},
  \bibinfo {author} {\bibfnamefont {A.}~\bibnamefont {Vishwanath}}, \ and\
  \bibinfo {author} {\bibfnamefont {S.~Y.}\ \bibnamefont {Savrasov}},\ }\href
  {\doibase 10.1103/PhysRevB.83.205101} {\bibfield  {journal} {\bibinfo
  {journal} {Phys. Rev. B}\ }\textbf {\bibinfo {volume} {83}},\ \bibinfo
  {pages} {205101} (\bibinfo {year} {2011})}\BibitemShut {NoStop}%
\bibitem [{\citenamefont {Weng}\ \emph {et~al.}(2015)\citenamefont {Weng},
  \citenamefont {Fang}, \citenamefont {Fang}, \citenamefont {Bernevig},\ and\
  \citenamefont {Dai}}]{wengWeyl2015}%
  \BibitemOpen
  \bibfield  {author} {\bibinfo {author} {\bibfnamefont {H.}~\bibnamefont
  {Weng}}, \bibinfo {author} {\bibfnamefont {C.}~\bibnamefont {Fang}}, \bibinfo
  {author} {\bibfnamefont {Z.}~\bibnamefont {Fang}}, \bibinfo {author}
  {\bibfnamefont {B.~A.}\ \bibnamefont {Bernevig}}, \ and\ \bibinfo {author}
  {\bibfnamefont {X.}~\bibnamefont {Dai}},\ }\href
  {https://link.aps.org/doi/10.1103/PhysRevX.5.011029} {\bibfield  {journal}
  {\bibinfo  {journal} {Phys. Rev. X}\ }\textbf {\bibinfo {volume} {5}}
  (\bibinfo {year} {2015})}\BibitemShut {NoStop}%
\bibitem [{\citenamefont {Soluyanov}\ \emph {et~al.}(2015)\citenamefont
  {Soluyanov}, \citenamefont {Gresch}, \citenamefont {Wang}, \citenamefont
  {Wu}, \citenamefont {Troyer}, \citenamefont {Dai},\ and\ \citenamefont
  {Bernevig}}]{soluyanov2015}%
  \BibitemOpen
  \bibfield  {author} {\bibinfo {author} {\bibfnamefont {A.~A.}\ \bibnamefont
  {Soluyanov}}, \bibinfo {author} {\bibfnamefont {D.}~\bibnamefont {Gresch}},
  \bibinfo {author} {\bibfnamefont {Z.}~\bibnamefont {Wang}}, \bibinfo {author}
  {\bibfnamefont {Q.}~\bibnamefont {Wu}}, \bibinfo {author} {\bibfnamefont
  {M.}~\bibnamefont {Troyer}}, \bibinfo {author} {\bibfnamefont
  {X.}~\bibnamefont {Dai}}, \ and\ \bibinfo {author} {\bibfnamefont {B.~A.}\
  \bibnamefont {Bernevig}},\ }\href {\doibase 10.1038/nature15768} {\bibfield
  {journal} {\bibinfo  {journal} {Nature}\ }\textbf {\bibinfo {volume} {527}},\
  \bibinfo {pages} {495} (\bibinfo {year} {2015})}\BibitemShut {NoStop}%
\bibitem [{\citenamefont {Kim}\ \emph {et~al.}(2009)\citenamefont {Kim},
  \citenamefont {Ohsumi}, \citenamefont {Komesu}, \citenamefont {Sakai},
  \citenamefont {Morita}, \citenamefont {Takagi},\ and\ \citenamefont
  {Arima}}]{sc1}%
  \BibitemOpen
  \bibfield  {author} {\bibinfo {author} {\bibfnamefont {B.}~\bibnamefont
  {Kim}}, \bibinfo {author} {\bibfnamefont {H.}~\bibnamefont {Ohsumi}},
  \bibinfo {author} {\bibfnamefont {T.}~\bibnamefont {Komesu}}, \bibinfo
  {author} {\bibfnamefont {S.}~\bibnamefont {Sakai}}, \bibinfo {author}
  {\bibfnamefont {T.}~\bibnamefont {Morita}}, \bibinfo {author} {\bibfnamefont
  {H.}~\bibnamefont {Takagi}}, \ and\ \bibinfo {author} {\bibfnamefont {T.-h.}\
  \bibnamefont {Arima}},\ }\href@noop {} {\bibfield  {journal} {\bibinfo
  {journal} {Science}\ }\textbf {\bibinfo {volume} {323}},\ \bibinfo {pages}
  {1329} (\bibinfo {year} {2009})}\BibitemShut {NoStop}%
\bibitem [{\citenamefont {Maeno}\ \emph {et~al.}(1994)\citenamefont {Maeno},
  \citenamefont {Hashimoto}, \citenamefont {Yoshida}, \citenamefont
  {Nishizaki}, \citenamefont {Fujita}, \citenamefont {Bednorz},\ and\
  \citenamefont {Lichtenberg}}]{sc2}%
  \BibitemOpen
  \bibfield  {author} {\bibinfo {author} {\bibfnamefont {Y.}~\bibnamefont
  {Maeno}}, \bibinfo {author} {\bibfnamefont {H.}~\bibnamefont {Hashimoto}},
  \bibinfo {author} {\bibfnamefont {K.}~\bibnamefont {Yoshida}}, \bibinfo
  {author} {\bibfnamefont {S.}~\bibnamefont {Nishizaki}}, \bibinfo {author}
  {\bibfnamefont {T.}~\bibnamefont {Fujita}}, \bibinfo {author} {\bibfnamefont
  {J.}~\bibnamefont {Bednorz}}, \ and\ \bibinfo {author} {\bibfnamefont
  {F.}~\bibnamefont {Lichtenberg}},\ }\href@noop {} {\bibfield  {journal}
  {\bibinfo  {journal} {nature}\ }\textbf {\bibinfo {volume} {372}},\ \bibinfo
  {pages} {532} (\bibinfo {year} {1994})}\BibitemShut {NoStop}%
\bibitem [{\citenamefont {Khaliullin}\ \emph {et~al.}(2004)\citenamefont
  {Khaliullin}, \citenamefont {Koshibae},\ and\ \citenamefont {Maekawa}}]{sc3}%
  \BibitemOpen
  \bibfield  {author} {\bibinfo {author} {\bibfnamefont {G.}~\bibnamefont
  {Khaliullin}}, \bibinfo {author} {\bibfnamefont {W.}~\bibnamefont
  {Koshibae}}, \ and\ \bibinfo {author} {\bibfnamefont {S.}~\bibnamefont
  {Maekawa}},\ }\href {\doibase 10.1103/PhysRevLett.93.176401} {\bibfield
  {journal} {\bibinfo  {journal} {Phys. Rev. Lett.}\ }\textbf {\bibinfo
  {volume} {93}},\ \bibinfo {pages} {176401} (\bibinfo {year}
  {2004})}\BibitemShut {NoStop}%
\bibitem [{\citenamefont {Kim}\ \emph {et~al.}(2008)\citenamefont {Kim},
  \citenamefont {Jin}, \citenamefont {Moon}, \citenamefont {Kim}, \citenamefont
  {Park}, \citenamefont {Leem}, \citenamefont {Yu}, \citenamefont {Noh},
  \citenamefont {Kim}, \citenamefont {Oh}, \citenamefont {Park}, \citenamefont
  {Durairaj}, \citenamefont {Cao},\ and\ \citenamefont {Rotenberg}}]{sc4}%
  \BibitemOpen
  \bibfield  {author} {\bibinfo {author} {\bibfnamefont {B.~J.}\ \bibnamefont
  {Kim}}, \bibinfo {author} {\bibfnamefont {H.}~\bibnamefont {Jin}}, \bibinfo
  {author} {\bibfnamefont {S.~J.}\ \bibnamefont {Moon}}, \bibinfo {author}
  {\bibfnamefont {J.-Y.}\ \bibnamefont {Kim}}, \bibinfo {author} {\bibfnamefont
  {B.-G.}\ \bibnamefont {Park}}, \bibinfo {author} {\bibfnamefont {C.~S.}\
  \bibnamefont {Leem}}, \bibinfo {author} {\bibfnamefont {J.}~\bibnamefont
  {Yu}}, \bibinfo {author} {\bibfnamefont {T.~W.}\ \bibnamefont {Noh}},
  \bibinfo {author} {\bibfnamefont {C.}~\bibnamefont {Kim}}, \bibinfo {author}
  {\bibfnamefont {S.-J.}\ \bibnamefont {Oh}}, \bibinfo {author} {\bibfnamefont
  {J.-H.}\ \bibnamefont {Park}}, \bibinfo {author} {\bibfnamefont
  {V.}~\bibnamefont {Durairaj}}, \bibinfo {author} {\bibfnamefont
  {G.}~\bibnamefont {Cao}}, \ and\ \bibinfo {author} {\bibfnamefont
  {E.}~\bibnamefont {Rotenberg}},\ }\href {\doibase
  10.1103/PhysRevLett.101.076402} {\bibfield  {journal} {\bibinfo  {journal}
  {Phys. Rev. Lett.}\ }\textbf {\bibinfo {volume} {101}},\ \bibinfo {pages}
  {076402} (\bibinfo {year} {2008})}\BibitemShut {NoStop}%
\bibitem [{\citenamefont {Khaliullin}(2005)}]{qsl1}%
  \BibitemOpen
  \bibfield  {author} {\bibinfo {author} {\bibfnamefont {G.}~\bibnamefont
  {Khaliullin}},\ }\href@noop {} {\bibfield  {journal} {\bibinfo  {journal}
  {Progress of Theoretical Physics Supplement}\ }\textbf {\bibinfo {volume}
  {160}},\ \bibinfo {pages} {155} (\bibinfo {year} {2005})}\BibitemShut
  {NoStop}%
\bibitem [{\citenamefont {Jackeli}\ and\ \citenamefont
  {Khaliullin}(2009)}]{qsl2}%
  \BibitemOpen
  \bibfield  {author} {\bibinfo {author} {\bibfnamefont {G.}~\bibnamefont
  {Jackeli}}\ and\ \bibinfo {author} {\bibfnamefont {G.}~\bibnamefont
  {Khaliullin}},\ }\href@noop {} {\bibfield  {journal} {\bibinfo  {journal}
  {Physical review letters}\ }\textbf {\bibinfo {volume} {102}},\ \bibinfo
  {pages} {017205} (\bibinfo {year} {2009})}\BibitemShut {NoStop}%
\bibitem [{\citenamefont {Zibouche}\ \emph {et~al.}(2014)\citenamefont
  {Zibouche}, \citenamefont {Kuc}, \citenamefont {Musfeldt},\ and\
  \citenamefont {Heine}}]{tmd}%
  \BibitemOpen
  \bibfield  {author} {\bibinfo {author} {\bibfnamefont {N.}~\bibnamefont
  {Zibouche}}, \bibinfo {author} {\bibfnamefont {A.}~\bibnamefont {Kuc}},
  \bibinfo {author} {\bibfnamefont {J.}~\bibnamefont {Musfeldt}}, \ and\
  \bibinfo {author} {\bibfnamefont {T.}~\bibnamefont {Heine}},\ }\href
  {\doibase https://doi.org/10.1002/andp.201400137} {\bibfield  {journal}
  {\bibinfo  {journal} {Annalen der Physik}\ }\textbf {\bibinfo {volume}
  {526}},\ \bibinfo {pages} {395} (\bibinfo {year} {2014})}\BibitemShut
  {NoStop}%
\bibitem [{\citenamefont {Georges}\ \emph {et~al.}(2013)\citenamefont
  {Georges}, \citenamefont {Medici},\ and\ \citenamefont {Mravlje}}]{hubbkan}%
  \BibitemOpen
  \bibfield  {author} {\bibinfo {author} {\bibfnamefont {A.}~\bibnamefont
  {Georges}}, \bibinfo {author} {\bibfnamefont {L.~d.}\ \bibnamefont {Medici}},
  \ and\ \bibinfo {author} {\bibfnamefont {J.}~\bibnamefont {Mravlje}},\
  }\href@noop {} {\bibfield  {journal} {\bibinfo  {journal} {Annu. Rev.
  Condens. Matter Phys.}\ }\textbf {\bibinfo {volume} {4}},\ \bibinfo {pages}
  {137} (\bibinfo {year} {2013})}\BibitemShut {NoStop}%
\bibitem [{\citenamefont {Kane}\ and\ \citenamefont {Mele}(2005)}]{kanem}%
  \BibitemOpen
  \bibfield  {author} {\bibinfo {author} {\bibfnamefont {C.~L.}\ \bibnamefont
  {Kane}}\ and\ \bibinfo {author} {\bibfnamefont {E.~J.}\ \bibnamefont
  {Mele}},\ }\href {\doibase 10.1103/PhysRevLett.95.226801} {\bibfield
  {journal} {\bibinfo  {journal} {Phys. Rev. Lett.}\ }\textbf {\bibinfo
  {volume} {95}},\ \bibinfo {pages} {226801} (\bibinfo {year}
  {2005})}\BibitemShut {NoStop}%
\bibitem [{\citenamefont {Pandey}\ \emph {et~al.}(2022)\citenamefont {Pandey},
  \citenamefont {Gu},\ and\ \citenamefont {Tiwari}}]{srrho}%
  \BibitemOpen
  \bibfield  {author} {\bibinfo {author} {\bibfnamefont {S.~K.}\ \bibnamefont
  {Pandey}}, \bibinfo {author} {\bibfnamefont {Q.}~\bibnamefont {Gu}}, \ and\
  \bibinfo {author} {\bibfnamefont {R.}~\bibnamefont {Tiwari}},\ }\href@noop {}
  {\bibfield  {journal} {\bibinfo  {journal} {arXiv preprint arXiv:2207.05045}\
  } (\bibinfo {year} {2022})}\BibitemShut {NoStop}%
\bibitem [{\citenamefont {Pandey}\ and\ \citenamefont {Feng}(2022)}]{coprb}%
  \BibitemOpen
  \bibfield  {author} {\bibinfo {author} {\bibfnamefont {S.~K.}\ \bibnamefont
  {Pandey}}\ and\ \bibinfo {author} {\bibfnamefont {J.}~\bibnamefont {Feng}},\
  }\href@noop {} {\bibfield  {journal} {\bibinfo  {journal} {arXiv preprint
  arXiv:2205.03836}\ } (\bibinfo {year} {2022})}\BibitemShut {NoStop}%
\bibitem [{\citenamefont {Adler}(1962)}]{crpa1}%
  \BibitemOpen
  \bibfield  {author} {\bibinfo {author} {\bibfnamefont {S.~L.}\ \bibnamefont
  {Adler}},\ }\href {\doibase 10.1103/PhysRev.126.413} {\bibfield  {journal}
  {\bibinfo  {journal} {Phys. Rev.}\ }\textbf {\bibinfo {volume} {126}},\
  \bibinfo {pages} {413} (\bibinfo {year} {1962})}\BibitemShut {NoStop}%
\bibitem [{\citenamefont {Wiser}(1963)}]{crpa2}%
  \BibitemOpen
  \bibfield  {author} {\bibinfo {author} {\bibfnamefont {N.}~\bibnamefont
  {Wiser}},\ }\href {\doibase 10.1103/PhysRev.129.62} {\bibfield  {journal}
  {\bibinfo  {journal} {Phys. Rev.}\ }\textbf {\bibinfo {volume} {129}},\
  \bibinfo {pages} {62} (\bibinfo {year} {1963})}\BibitemShut {NoStop}%
\bibitem [{\citenamefont {Aryasetiawan}\ \emph {et~al.}(2004)\citenamefont
  {Aryasetiawan}, \citenamefont {Imada}, \citenamefont {Georges}, \citenamefont
  {Kotliar}, \citenamefont {Biermann},\ and\ \citenamefont
  {Lichtenstein}}]{crpa3}%
  \BibitemOpen
  \bibfield  {author} {\bibinfo {author} {\bibfnamefont {F.}~\bibnamefont
  {Aryasetiawan}}, \bibinfo {author} {\bibfnamefont {M.}~\bibnamefont {Imada}},
  \bibinfo {author} {\bibfnamefont {A.}~\bibnamefont {Georges}}, \bibinfo
  {author} {\bibfnamefont {G.}~\bibnamefont {Kotliar}}, \bibinfo {author}
  {\bibfnamefont {S.}~\bibnamefont {Biermann}}, \ and\ \bibinfo {author}
  {\bibfnamefont {A.~I.}\ \bibnamefont {Lichtenstein}},\ }\href {\doibase
  10.1103/PhysRevB.70.195104} {\bibfield  {journal} {\bibinfo  {journal} {Phys.
  Rev. B}\ }\textbf {\bibinfo {volume} {70}},\ \bibinfo {pages} {195104}
  (\bibinfo {year} {2004})}\BibitemShut {NoStop}%
\bibitem [{\citenamefont {Marzari}\ \emph {et~al.}(2012)\citenamefont
  {Marzari}, \citenamefont {Mostofi}, \citenamefont {Yates}, \citenamefont
  {Souza},\ and\ \citenamefont {Vanderbilt}}]{Marzari2012}%
  \BibitemOpen
  \bibfield  {author} {\bibinfo {author} {\bibfnamefont {N.}~\bibnamefont
  {Marzari}}, \bibinfo {author} {\bibfnamefont {A.~A.}\ \bibnamefont
  {Mostofi}}, \bibinfo {author} {\bibfnamefont {J.~R.}\ \bibnamefont {Yates}},
  \bibinfo {author} {\bibfnamefont {I.}~\bibnamefont {Souza}}, \ and\ \bibinfo
  {author} {\bibfnamefont {D.}~\bibnamefont {Vanderbilt}},\ }\href {\doibase
  10.1103/RevModPhys.84.1419} {\bibfield  {journal} {\bibinfo  {journal}
  {Reviews of Modern Physics}\ }\textbf {\bibinfo {volume} {84}},\ \bibinfo
  {pages} {1419} (\bibinfo {year} {2012})}\BibitemShut {NoStop}%
\bibitem [{\citenamefont {Jha}\ and\ \citenamefont {Heine}(2022)}]{soc1}%
  \BibitemOpen
  \bibfield  {author} {\bibinfo {author} {\bibfnamefont {G.}~\bibnamefont
  {Jha}}\ and\ \bibinfo {author} {\bibfnamefont {T.}~\bibnamefont {Heine}},\
  }\href {\doibase 10.1021/acs.jctc.2c00376} {\bibfield  {journal} {\bibinfo
  {journal} {Journal of Chemical Theory and Computation}\ }\textbf {\bibinfo
  {volume} {18}},\ \bibinfo {pages} {4472} (\bibinfo {year} {2022})},\ \bibinfo
  {note} {pMID: 35737969}\BibitemShut {NoStop}%
\bibitem [{\citenamefont {Cuadrado}\ \emph {et~al.}(2021)\citenamefont
  {Cuadrado}, \citenamefont {Robles}, \citenamefont {Garc\'{\i}a},
  \citenamefont {Pruneda}, \citenamefont {Ordej\'on}, \citenamefont {Ferrer},\
  and\ \citenamefont {Cerd\'a}}]{soc2}%
  \BibitemOpen
  \bibfield  {author} {\bibinfo {author} {\bibfnamefont {R.}~\bibnamefont
  {Cuadrado}}, \bibinfo {author} {\bibfnamefont {R.}~\bibnamefont {Robles}},
  \bibinfo {author} {\bibfnamefont {A.}~\bibnamefont {Garc\'{\i}a}}, \bibinfo
  {author} {\bibfnamefont {M.}~\bibnamefont {Pruneda}}, \bibinfo {author}
  {\bibfnamefont {P.}~\bibnamefont {Ordej\'on}}, \bibinfo {author}
  {\bibfnamefont {J.}~\bibnamefont {Ferrer}}, \ and\ \bibinfo {author}
  {\bibfnamefont {J.~I.}\ \bibnamefont {Cerd\'a}},\ }\href {\doibase
  10.1103/PhysRevB.104.195104} {\bibfield  {journal} {\bibinfo  {journal}
  {Phys. Rev. B}\ }\textbf {\bibinfo {volume} {104}},\ \bibinfo {pages}
  {195104} (\bibinfo {year} {2021})}\BibitemShut {NoStop}%
\bibitem [{\citenamefont {Kurita}\ and\ \citenamefont
  {Koretsune}(2020)}]{soc3}%
  \BibitemOpen
  \bibfield  {author} {\bibinfo {author} {\bibfnamefont {K.}~\bibnamefont
  {Kurita}}\ and\ \bibinfo {author} {\bibfnamefont {T.}~\bibnamefont
  {Koretsune}},\ }\href {\doibase 10.1103/PhysRevB.102.045109} {\bibfield
  {journal} {\bibinfo  {journal} {Phys. Rev. B}\ }\textbf {\bibinfo {volume}
  {102}},\ \bibinfo {pages} {045109} (\bibinfo {year} {2020})}\BibitemShut
  {NoStop}%
\bibitem [{\citenamefont {Vijayakumar}\ and\ \citenamefont
  {Gopinathan}(1996)}]{soc4}%
  \BibitemOpen
  \bibfield  {author} {\bibinfo {author} {\bibfnamefont {M.}~\bibnamefont
  {Vijayakumar}}\ and\ \bibinfo {author} {\bibfnamefont {M.}~\bibnamefont
  {Gopinathan}},\ }\href {\doibase
  https://doi.org/10.1016/0166-1280(95)04297-0} {\bibfield  {journal} {\bibinfo
   {journal} {Journal of Molecular Structure: THEOCHEM}\ }\textbf {\bibinfo
  {volume} {361}},\ \bibinfo {pages} {15} (\bibinfo {year} {1996})},\ \bibinfo
  {note} {theoretical Chemistry in India}\BibitemShut {NoStop}%
\bibitem [{\citenamefont {Gu}(2020)}]{qq}%
  \BibitemOpen
  \bibfield  {author} {\bibinfo {author} {\bibfnamefont {Q.}~\bibnamefont
  {Gu}},\ }\href@noop {} {\enquote {\bibinfo {title} {{TBSOC}},}\ }\bibinfo
  {howpublished} {\url{https://github.com/qqgu/TBSOC}} (\bibinfo {year}
  {2020})\BibitemShut {NoStop}%
\bibitem [{\citenamefont {Nelder}\ and\ \citenamefont
  {Mead}(1965)}]{NelderMead1965}%
  \BibitemOpen
  \bibfield  {author} {\bibinfo {author} {\bibfnamefont {J.~A.}\ \bibnamefont
  {Nelder}}\ and\ \bibinfo {author} {\bibfnamefont {R.}~\bibnamefont {Mead}},\
  }\href {\doibase 10.1093/comjnl/7.4.308} {\bibfield  {journal} {\bibinfo
  {journal} {The Computer Journal}\ }\textbf {\bibinfo {volume} {7}},\ \bibinfo
  {pages} {308} (\bibinfo {year} {1965})}\BibitemShut {NoStop}%
\bibitem [{\citenamefont {Gao}\ and\ \citenamefont {Han}(2012)}]{Gao2012}%
  \BibitemOpen
  \bibfield  {author} {\bibinfo {author} {\bibfnamefont {F.}~\bibnamefont
  {Gao}}\ and\ \bibinfo {author} {\bibfnamefont {L.}~\bibnamefont {Han}},\
  }\href {\doibase 10.1007/s10589-010-9329-3} {\bibfield  {journal} {\bibinfo
  {journal} {Computational Optimization and Applications}\ }\textbf {\bibinfo
  {volume} {51}},\ \bibinfo {pages} {259} (\bibinfo {year} {2012})}\BibitemShut
  {NoStop}%
\bibitem [{\citenamefont {Mostofi}\ \emph {et~al.}(2008)\citenamefont
  {Mostofi}, \citenamefont {Yates}, \citenamefont {Lee}, \citenamefont {Souza},
  \citenamefont {Vanderbilt},\ and\ \citenamefont {Marzari}}]{wannier90}%
  \BibitemOpen
  \bibfield  {author} {\bibinfo {author} {\bibfnamefont {A.~A.}\ \bibnamefont
  {Mostofi}}, \bibinfo {author} {\bibfnamefont {J.~R.}\ \bibnamefont {Yates}},
  \bibinfo {author} {\bibfnamefont {Y.-S.}\ \bibnamefont {Lee}}, \bibinfo
  {author} {\bibfnamefont {I.}~\bibnamefont {Souza}}, \bibinfo {author}
  {\bibfnamefont {D.}~\bibnamefont {Vanderbilt}}, \ and\ \bibinfo {author}
  {\bibfnamefont {N.}~\bibnamefont {Marzari}},\ }\href {\doibase
  10.1016/j.cpc.2007.11.016} {\bibfield  {journal} {\bibinfo  {journal}
  {Comput. Phys. Commun.}\ }\textbf {\bibinfo {volume} {178}},\ \bibinfo
  {pages} {685 } (\bibinfo {year} {2008})}\BibitemShut {NoStop}%
\bibitem [{\citenamefont {Kresse}\ and\ \citenamefont
  {Furthm\"uller}(1996)}]{Kresse}%
  \BibitemOpen
  \bibfield  {author} {\bibinfo {author} {\bibfnamefont {G.}~\bibnamefont
  {Kresse}}\ and\ \bibinfo {author} {\bibfnamefont {J.}~\bibnamefont
  {Furthm\"uller}},\ }\href {\doibase 10.1103/PhysRevB.54.11169} {\bibfield
  {journal} {\bibinfo  {journal} {Phys. Rev. B}\ }\textbf {\bibinfo {volume}
  {54}},\ \bibinfo {pages} {11169} (\bibinfo {year} {1996})}\BibitemShut
  {NoStop}%
\bibitem [{\citenamefont {Kresse}\ and\ \citenamefont {Joubert}(1999)}]{paw1}%
  \BibitemOpen
  \bibfield  {author} {\bibinfo {author} {\bibfnamefont {G.}~\bibnamefont
  {Kresse}}\ and\ \bibinfo {author} {\bibfnamefont {D.}~\bibnamefont
  {Joubert}},\ }\href {\doibase 10.1103/PhysRevB.59.1758} {\bibfield  {journal}
  {\bibinfo  {journal} {Phys. Rev. B}\ }\textbf {\bibinfo {volume} {59}},\
  \bibinfo {pages} {1758} (\bibinfo {year} {1999})}\BibitemShut {NoStop}%
\bibitem [{\citenamefont {Bl\"ochl}(1994)}]{paw2}%
  \BibitemOpen
  \bibfield  {author} {\bibinfo {author} {\bibfnamefont {P.~E.}\ \bibnamefont
  {Bl\"ochl}},\ }\href {\doibase 10.1103/PhysRevB.50.17953} {\bibfield
  {journal} {\bibinfo  {journal} {Phys. Rev. B}\ }\textbf {\bibinfo {volume}
  {50}},\ \bibinfo {pages} {17953} (\bibinfo {year} {1994})}\BibitemShut
  {NoStop}%
\bibitem [{\citenamefont {Perdew}\ \emph {et~al.}(1996)\citenamefont {Perdew},
  \citenamefont {Burke},\ and\ \citenamefont {Ernzerhof}}]{PBE}%
  \BibitemOpen
  \bibfield  {author} {\bibinfo {author} {\bibfnamefont {J.~P.}\ \bibnamefont
  {Perdew}}, \bibinfo {author} {\bibfnamefont {K.}~\bibnamefont {Burke}}, \
  and\ \bibinfo {author} {\bibfnamefont {M.}~\bibnamefont {Ernzerhof}},\ }\href
  {\doibase 10.1103/PhysRevLett.77.3865} {\bibfield  {journal} {\bibinfo
  {journal} {Phys. Rev. Lett.}\ }\textbf {\bibinfo {volume} {77}},\ \bibinfo
  {pages} {3865} (\bibinfo {year} {1996})}\BibitemShut {NoStop}%
\bibitem [{\citenamefont {Winter}\ \emph {et~al.}(2016)\citenamefont {Winter},
  \citenamefont {Li}, \citenamefont {Jeschke},\ and\ \citenamefont
  {Valent\'{\i}}}]{winterprb}%
  \BibitemOpen
  \bibfield  {author} {\bibinfo {author} {\bibfnamefont {S.~M.}\ \bibnamefont
  {Winter}}, \bibinfo {author} {\bibfnamefont {Y.}~\bibnamefont {Li}}, \bibinfo
  {author} {\bibfnamefont {H.~O.}\ \bibnamefont {Jeschke}}, \ and\ \bibinfo
  {author} {\bibfnamefont {R.}~\bibnamefont {Valent\'{\i}}},\ }\href {\doibase
  10.1103/PhysRevB.93.214431} {\bibfield  {journal} {\bibinfo  {journal} {Phys.
  Rev. B}\ }\textbf {\bibinfo {volume} {93}},\ \bibinfo {pages} {214431}
  (\bibinfo {year} {2016})}\BibitemShut {NoStop}%
\bibitem [{\citenamefont {Johnson}\ \emph {et~al.}(2015)\citenamefont
  {Johnson}, \citenamefont {Williams}, \citenamefont {Haghighirad},
  \citenamefont {Singleton}, \citenamefont {Zapf}, \citenamefont {Manuel},
  \citenamefont {Mazin}, \citenamefont {Li}, \citenamefont {Jeschke},
  \citenamefont {Valent\'{\i}},\ and\ \citenamefont {Coldea}}]{rcl_exp1}%
  \BibitemOpen
  \bibfield  {author} {\bibinfo {author} {\bibfnamefont {R.~D.}\ \bibnamefont
  {Johnson}}, \bibinfo {author} {\bibfnamefont {S.~C.}\ \bibnamefont
  {Williams}}, \bibinfo {author} {\bibfnamefont {A.~A.}\ \bibnamefont
  {Haghighirad}}, \bibinfo {author} {\bibfnamefont {J.}~\bibnamefont
  {Singleton}}, \bibinfo {author} {\bibfnamefont {V.}~\bibnamefont {Zapf}},
  \bibinfo {author} {\bibfnamefont {P.}~\bibnamefont {Manuel}}, \bibinfo
  {author} {\bibfnamefont {I.~I.}\ \bibnamefont {Mazin}}, \bibinfo {author}
  {\bibfnamefont {Y.}~\bibnamefont {Li}}, \bibinfo {author} {\bibfnamefont
  {H.~O.}\ \bibnamefont {Jeschke}}, \bibinfo {author} {\bibfnamefont
  {R.}~\bibnamefont {Valent\'{\i}}}, \ and\ \bibinfo {author} {\bibfnamefont
  {R.}~\bibnamefont {Coldea}},\ }\href {\doibase 10.1103/PhysRevB.92.235119}
  {\bibfield  {journal} {\bibinfo  {journal} {Phys. Rev. B}\ }\textbf {\bibinfo
  {volume} {92}},\ \bibinfo {pages} {235119} (\bibinfo {year}
  {2015})}\BibitemShut {NoStop}%
\bibitem [{\citenamefont {Cao}\ \emph {et~al.}(2016)\citenamefont {Cao},
  \citenamefont {Banerjee}, \citenamefont {Yan}, \citenamefont {Bridges},
  \citenamefont {Lumsden}, \citenamefont {Mandrus}, \citenamefont {Tennant},
  \citenamefont {Chakoumakos},\ and\ \citenamefont {Nagler}}]{rcl_exp2}%
  \BibitemOpen
  \bibfield  {author} {\bibinfo {author} {\bibfnamefont {H.~B.}\ \bibnamefont
  {Cao}}, \bibinfo {author} {\bibfnamefont {A.}~\bibnamefont {Banerjee}},
  \bibinfo {author} {\bibfnamefont {J.-Q.}\ \bibnamefont {Yan}}, \bibinfo
  {author} {\bibfnamefont {C.~A.}\ \bibnamefont {Bridges}}, \bibinfo {author}
  {\bibfnamefont {M.~D.}\ \bibnamefont {Lumsden}}, \bibinfo {author}
  {\bibfnamefont {D.~G.}\ \bibnamefont {Mandrus}}, \bibinfo {author}
  {\bibfnamefont {D.~A.}\ \bibnamefont {Tennant}}, \bibinfo {author}
  {\bibfnamefont {B.~C.}\ \bibnamefont {Chakoumakos}}, \ and\ \bibinfo {author}
  {\bibfnamefont {S.~E.}\ \bibnamefont {Nagler}},\ }\href {\doibase
  10.1103/PhysRevB.93.134423} {\bibfield  {journal} {\bibinfo  {journal} {Phys.
  Rev. B}\ }\textbf {\bibinfo {volume} {93}},\ \bibinfo {pages} {134423}
  (\bibinfo {year} {2016})}\BibitemShut {NoStop}%
\bibitem [{\citenamefont {Brodersen}\ \emph {et~al.}(1968)\citenamefont
  {Brodersen}, \citenamefont {Thiele}, \citenamefont {Ohnsorge}, \citenamefont
  {Recke},\ and\ \citenamefont {Moers}}]{rcl_exp3}%
  \BibitemOpen
  \bibfield  {author} {\bibinfo {author} {\bibfnamefont {K.}~\bibnamefont
  {Brodersen}}, \bibinfo {author} {\bibfnamefont {G.}~\bibnamefont {Thiele}},
  \bibinfo {author} {\bibfnamefont {H.}~\bibnamefont {Ohnsorge}}, \bibinfo
  {author} {\bibfnamefont {I.}~\bibnamefont {Recke}}, \ and\ \bibinfo {author}
  {\bibfnamefont {F.}~\bibnamefont {Moers}},\ }\href {\doibase
  https://doi.org/10.1016/0022-5088(68)90194-X} {\bibfield  {journal} {\bibinfo
   {journal} {Journal of the Less Common Metals}\ }\textbf {\bibinfo {volume}
  {15}},\ \bibinfo {pages} {347} (\bibinfo {year} {1968})}\BibitemShut
  {NoStop}%
\bibitem [{\citenamefont {Plumb}\ \emph {et~al.}(2014)\citenamefont {Plumb},
  \citenamefont {Clancy}, \citenamefont {Sandilands}, \citenamefont {Shankar},
  \citenamefont {Hu}, \citenamefont {Burch}, \citenamefont {Kee},\ and\
  \citenamefont {Kim}}]{rcl_soc_exp1}%
  \BibitemOpen
  \bibfield  {author} {\bibinfo {author} {\bibfnamefont {K.~W.}\ \bibnamefont
  {Plumb}}, \bibinfo {author} {\bibfnamefont {J.~P.}\ \bibnamefont {Clancy}},
  \bibinfo {author} {\bibfnamefont {L.~J.}\ \bibnamefont {Sandilands}},
  \bibinfo {author} {\bibfnamefont {V.~V.}\ \bibnamefont {Shankar}}, \bibinfo
  {author} {\bibfnamefont {Y.~F.}\ \bibnamefont {Hu}}, \bibinfo {author}
  {\bibfnamefont {K.~S.}\ \bibnamefont {Burch}}, \bibinfo {author}
  {\bibfnamefont {H.-Y.}\ \bibnamefont {Kee}}, \ and\ \bibinfo {author}
  {\bibfnamefont {Y.-J.}\ \bibnamefont {Kim}},\ }\href {\doibase
  10.1103/PhysRevB.90.041112} {\bibfield  {journal} {\bibinfo  {journal} {Phys.
  Rev. B}\ }\textbf {\bibinfo {volume} {90}},\ \bibinfo {pages} {041112}
  (\bibinfo {year} {2014})}\BibitemShut {NoStop}%
\bibitem [{\citenamefont {Sandilands}\ \emph {et~al.}(2016)\citenamefont
  {Sandilands}, \citenamefont {Tian}, \citenamefont {Reijnders}, \citenamefont
  {Kim}, \citenamefont {Plumb}, \citenamefont {Kim}, \citenamefont {Kee},\ and\
  \citenamefont {Burch}}]{rcl_soc_exp2}%
  \BibitemOpen
  \bibfield  {author} {\bibinfo {author} {\bibfnamefont {L.~J.}\ \bibnamefont
  {Sandilands}}, \bibinfo {author} {\bibfnamefont {Y.}~\bibnamefont {Tian}},
  \bibinfo {author} {\bibfnamefont {A.~A.}\ \bibnamefont {Reijnders}}, \bibinfo
  {author} {\bibfnamefont {H.-S.}\ \bibnamefont {Kim}}, \bibinfo {author}
  {\bibfnamefont {K.~W.}\ \bibnamefont {Plumb}}, \bibinfo {author}
  {\bibfnamefont {Y.-J.}\ \bibnamefont {Kim}}, \bibinfo {author} {\bibfnamefont
  {H.-Y.}\ \bibnamefont {Kee}}, \ and\ \bibinfo {author} {\bibfnamefont
  {K.~S.}\ \bibnamefont {Burch}},\ }\href {\doibase 10.1103/PhysRevB.93.075144}
  {\bibfield  {journal} {\bibinfo  {journal} {Phys. Rev. B}\ }\textbf {\bibinfo
  {volume} {93}},\ \bibinfo {pages} {075144} (\bibinfo {year}
  {2016})}\BibitemShut {NoStop}%
\bibitem [{\citenamefont {Songvilay}\ \emph {et~al.}(2020)\citenamefont
  {Songvilay}, \citenamefont {Robert}, \citenamefont {Petit}, \citenamefont
  {Rodriguez-Rivera}, \citenamefont {Ratcliff}, \citenamefont {Damay},
  \citenamefont {Bal\'edent}, \citenamefont {Jim\'enez-Ruiz}, \citenamefont
  {Lejay}, \citenamefont {Pachoud}, \citenamefont {Hadj-Azzem}, \citenamefont
  {Simonet},\ and\ \citenamefont {Stock}}]{co_exp1}%
  \BibitemOpen
  \bibfield  {author} {\bibinfo {author} {\bibfnamefont {M.}~\bibnamefont
  {Songvilay}}, \bibinfo {author} {\bibfnamefont {J.}~\bibnamefont {Robert}},
  \bibinfo {author} {\bibfnamefont {S.}~\bibnamefont {Petit}}, \bibinfo
  {author} {\bibfnamefont {J.~A.}\ \bibnamefont {Rodriguez-Rivera}}, \bibinfo
  {author} {\bibfnamefont {W.~D.}\ \bibnamefont {Ratcliff}}, \bibinfo {author}
  {\bibfnamefont {F.}~\bibnamefont {Damay}}, \bibinfo {author} {\bibfnamefont
  {V.}~\bibnamefont {Bal\'edent}}, \bibinfo {author} {\bibfnamefont
  {M.}~\bibnamefont {Jim\'enez-Ruiz}}, \bibinfo {author} {\bibfnamefont
  {P.}~\bibnamefont {Lejay}}, \bibinfo {author} {\bibfnamefont
  {E.}~\bibnamefont {Pachoud}}, \bibinfo {author} {\bibfnamefont
  {A.}~\bibnamefont {Hadj-Azzem}}, \bibinfo {author} {\bibfnamefont
  {V.}~\bibnamefont {Simonet}}, \ and\ \bibinfo {author} {\bibfnamefont
  {C.}~\bibnamefont {Stock}},\ }\href {\doibase 10.1103/PhysRevB.102.224429}
  {\bibfield  {journal} {\bibinfo  {journal} {Phys. Rev. B}\ }\textbf {\bibinfo
  {volume} {102}},\ \bibinfo {pages} {224429} (\bibinfo {year}
  {2020})}\BibitemShut {NoStop}%
\bibitem [{\citenamefont {Viciu}\ \emph {et~al.}(2007)\citenamefont {Viciu},
  \citenamefont {Huang}, \citenamefont {Morosan}, \citenamefont {Zandbergen},
  \citenamefont {Greenbaum}, \citenamefont {McQueen},\ and\ \citenamefont
  {Cava}}]{co_exp2}%
  \BibitemOpen
  \bibfield  {author} {\bibinfo {author} {\bibfnamefont {L.}~\bibnamefont
  {Viciu}}, \bibinfo {author} {\bibfnamefont {Q.}~\bibnamefont {Huang}},
  \bibinfo {author} {\bibfnamefont {E.}~\bibnamefont {Morosan}}, \bibinfo
  {author} {\bibfnamefont {H.}~\bibnamefont {Zandbergen}}, \bibinfo {author}
  {\bibfnamefont {N.}~\bibnamefont {Greenbaum}}, \bibinfo {author}
  {\bibfnamefont {T.}~\bibnamefont {McQueen}}, \ and\ \bibinfo {author}
  {\bibfnamefont {R.}~\bibnamefont {Cava}},\ }\href {\doibase
  https://doi.org/10.1016/j.jssc.2007.01.002} {\bibfield  {journal} {\bibinfo
  {journal} {Journal of Solid State Chemistry}\ }\textbf {\bibinfo {volume}
  {180}},\ \bibinfo {pages} {1060} (\bibinfo {year} {2007})}\BibitemShut
  {NoStop}%
\bibitem [{\citenamefont {Chen}\ \emph {et~al.}(2021)\citenamefont {Chen},
  \citenamefont {Li}, \citenamefont {Hu}, \citenamefont {Hu}, \citenamefont
  {Yue}, \citenamefont {Sutarto}, \citenamefont {He}, \citenamefont {Iida},
  \citenamefont {Kamazawa}, \citenamefont {Yu}, \citenamefont {Lin},\ and\
  \citenamefont {Li}}]{co_exp3}%
  \BibitemOpen
  \bibfield  {author} {\bibinfo {author} {\bibfnamefont {W.}~\bibnamefont
  {Chen}}, \bibinfo {author} {\bibfnamefont {X.}~\bibnamefont {Li}}, \bibinfo
  {author} {\bibfnamefont {Z.}~\bibnamefont {Hu}}, \bibinfo {author}
  {\bibfnamefont {Z.}~\bibnamefont {Hu}}, \bibinfo {author} {\bibfnamefont
  {L.}~\bibnamefont {Yue}}, \bibinfo {author} {\bibfnamefont {R.}~\bibnamefont
  {Sutarto}}, \bibinfo {author} {\bibfnamefont {F.}~\bibnamefont {He}},
  \bibinfo {author} {\bibfnamefont {K.}~\bibnamefont {Iida}}, \bibinfo {author}
  {\bibfnamefont {K.}~\bibnamefont {Kamazawa}}, \bibinfo {author}
  {\bibfnamefont {W.}~\bibnamefont {Yu}}, \bibinfo {author} {\bibfnamefont
  {X.}~\bibnamefont {Lin}}, \ and\ \bibinfo {author} {\bibfnamefont
  {Y.}~\bibnamefont {Li}},\ }\href {\doibase 10.1103/PhysRevB.103.L180404}
  {\bibfield  {journal} {\bibinfo  {journal} {Phys. Rev. B}\ }\textbf {\bibinfo
  {volume} {103}},\ \bibinfo {pages} {L180404} (\bibinfo {year}
  {2021})}\BibitemShut {NoStop}%
\bibitem [{\citenamefont {Xiao}\ \emph {et~al.}(2019)\citenamefont {Xiao},
  \citenamefont {Xia}, \citenamefont {Zhang}, \citenamefont {Yue},
  \citenamefont {Huang}, \citenamefont {Zhang}, \citenamefont {Yang},
  \citenamefont {Song}, \citenamefont {Wei}, \citenamefont {Deng} \emph
  {et~al.}}]{co_exp4}%
  \BibitemOpen
  \bibfield  {author} {\bibinfo {author} {\bibfnamefont {G.}~\bibnamefont
  {Xiao}}, \bibinfo {author} {\bibfnamefont {Z.}~\bibnamefont {Xia}}, \bibinfo
  {author} {\bibfnamefont {W.}~\bibnamefont {Zhang}}, \bibinfo {author}
  {\bibfnamefont {X.}~\bibnamefont {Yue}}, \bibinfo {author} {\bibfnamefont
  {S.}~\bibnamefont {Huang}}, \bibinfo {author} {\bibfnamefont
  {X.}~\bibnamefont {Zhang}}, \bibinfo {author} {\bibfnamefont
  {F.}~\bibnamefont {Yang}}, \bibinfo {author} {\bibfnamefont {Y.}~\bibnamefont
  {Song}}, \bibinfo {author} {\bibfnamefont {M.}~\bibnamefont {Wei}}, \bibinfo
  {author} {\bibfnamefont {H.}~\bibnamefont {Deng}},  \emph {et~al.},\
  }\href@noop {} {\bibfield  {journal} {\bibinfo  {journal} {Crystal Growth \&
  Design}\ }\textbf {\bibinfo {volume} {19}},\ \bibinfo {pages} {2658}
  (\bibinfo {year} {2019})}\BibitemShut {NoStop}%
\bibitem [{\citenamefont {Lefran\ifmmode~\mbox{\c{c}}\else \c{c}\fi{}ois}\
  \emph {et~al.}(2016)\citenamefont {Lefran\ifmmode~\mbox{\c{c}}\else
  \c{c}\fi{}ois}, \citenamefont {Songvilay}, \citenamefont {Robert},
  \citenamefont {Nataf}, \citenamefont {Jordan}, \citenamefont {Chaix},
  \citenamefont {Colin}, \citenamefont {Lejay}, \citenamefont {Hadj-Azzem},
  \citenamefont {Ballou},\ and\ \citenamefont {Simonet}}]{co_exp5}%
  \BibitemOpen
  \bibfield  {author} {\bibinfo {author} {\bibfnamefont {E.}~\bibnamefont
  {Lefran\ifmmode~\mbox{\c{c}}\else \c{c}\fi{}ois}}, \bibinfo {author}
  {\bibfnamefont {M.}~\bibnamefont {Songvilay}}, \bibinfo {author}
  {\bibfnamefont {J.}~\bibnamefont {Robert}}, \bibinfo {author} {\bibfnamefont
  {G.}~\bibnamefont {Nataf}}, \bibinfo {author} {\bibfnamefont
  {E.}~\bibnamefont {Jordan}}, \bibinfo {author} {\bibfnamefont
  {L.}~\bibnamefont {Chaix}}, \bibinfo {author} {\bibfnamefont {C.~V.}\
  \bibnamefont {Colin}}, \bibinfo {author} {\bibfnamefont {P.}~\bibnamefont
  {Lejay}}, \bibinfo {author} {\bibfnamefont {A.}~\bibnamefont {Hadj-Azzem}},
  \bibinfo {author} {\bibfnamefont {R.}~\bibnamefont {Ballou}}, \ and\ \bibinfo
  {author} {\bibfnamefont {V.}~\bibnamefont {Simonet}},\ }\href {\doibase
  10.1103/PhysRevB.94.214416} {\bibfield  {journal} {\bibinfo  {journal} {Phys.
  Rev. B}\ }\textbf {\bibinfo {volume} {94}},\ \bibinfo {pages} {214416}
  (\bibinfo {year} {2016})}\BibitemShut {NoStop}%
\bibitem [{\citenamefont {Bera}\ \emph {et~al.}(2017)\citenamefont {Bera},
  \citenamefont {Yusuf}, \citenamefont {Kumar},\ and\ \citenamefont
  {Ritter}}]{co_exp6}%
  \BibitemOpen
  \bibfield  {author} {\bibinfo {author} {\bibfnamefont {A.~K.}\ \bibnamefont
  {Bera}}, \bibinfo {author} {\bibfnamefont {S.~M.}\ \bibnamefont {Yusuf}},
  \bibinfo {author} {\bibfnamefont {A.}~\bibnamefont {Kumar}}, \ and\ \bibinfo
  {author} {\bibfnamefont {C.}~\bibnamefont {Ritter}},\ }\href {\doibase
  10.1103/PhysRevB.95.094424} {\bibfield  {journal} {\bibinfo  {journal} {Phys.
  Rev. B}\ }\textbf {\bibinfo {volume} {95}},\ \bibinfo {pages} {094424}
  (\bibinfo {year} {2017})}\BibitemShut {NoStop}%
\bibitem [{\citenamefont {Singh}\ \emph {et~al.}(2012)\citenamefont {Singh},
  \citenamefont {Manni}, \citenamefont {Reuther}, \citenamefont {Berlijn},
  \citenamefont {Thomale}, \citenamefont {Ku}, \citenamefont {Trebst},\ and\
  \citenamefont {Gegenwart}}]{ir_exp1}%
  \BibitemOpen
  \bibfield  {author} {\bibinfo {author} {\bibfnamefont {Y.}~\bibnamefont
  {Singh}}, \bibinfo {author} {\bibfnamefont {S.}~\bibnamefont {Manni}},
  \bibinfo {author} {\bibfnamefont {J.}~\bibnamefont {Reuther}}, \bibinfo
  {author} {\bibfnamefont {T.}~\bibnamefont {Berlijn}}, \bibinfo {author}
  {\bibfnamefont {R.}~\bibnamefont {Thomale}}, \bibinfo {author} {\bibfnamefont
  {W.}~\bibnamefont {Ku}}, \bibinfo {author} {\bibfnamefont {S.}~\bibnamefont
  {Trebst}}, \ and\ \bibinfo {author} {\bibfnamefont {P.}~\bibnamefont
  {Gegenwart}},\ }\href {\doibase 10.1103/PhysRevLett.108.127203} {\bibfield
  {journal} {\bibinfo  {journal} {Phys. Rev. Lett.}\ }\textbf {\bibinfo
  {volume} {108}},\ \bibinfo {pages} {127203} (\bibinfo {year}
  {2012})}\BibitemShut {NoStop}%
\bibitem [{\citenamefont {Singh}\ and\ \citenamefont
  {Gegenwart}(2010)}]{ir_exp2}%
  \BibitemOpen
  \bibfield  {author} {\bibinfo {author} {\bibfnamefont {Y.}~\bibnamefont
  {Singh}}\ and\ \bibinfo {author} {\bibfnamefont {P.}~\bibnamefont
  {Gegenwart}},\ }\href {\doibase 10.1103/PhysRevB.82.064412} {\bibfield
  {journal} {\bibinfo  {journal} {Phys. Rev. B}\ }\textbf {\bibinfo {volume}
  {82}},\ \bibinfo {pages} {064412} (\bibinfo {year} {2010})}\BibitemShut
  {NoStop}%
\bibitem [{\citenamefont {Choi}\ \emph {et~al.}(2012)\citenamefont {Choi},
  \citenamefont {Coldea}, \citenamefont {Kolmogorov}, \citenamefont
  {Lancaster}, \citenamefont {Mazin}, \citenamefont {Blundell}, \citenamefont
  {Radaelli}, \citenamefont {Singh}, \citenamefont {Gegenwart}, \citenamefont
  {Choi}, \citenamefont {Cheong}, \citenamefont {Baker}, \citenamefont
  {Stock},\ and\ \citenamefont {Taylor}}]{ir_exp3}%
  \BibitemOpen
  \bibfield  {author} {\bibinfo {author} {\bibfnamefont {S.~K.}\ \bibnamefont
  {Choi}}, \bibinfo {author} {\bibfnamefont {R.}~\bibnamefont {Coldea}},
  \bibinfo {author} {\bibfnamefont {A.~N.}\ \bibnamefont {Kolmogorov}},
  \bibinfo {author} {\bibfnamefont {T.}~\bibnamefont {Lancaster}}, \bibinfo
  {author} {\bibfnamefont {I.~I.}\ \bibnamefont {Mazin}}, \bibinfo {author}
  {\bibfnamefont {S.~J.}\ \bibnamefont {Blundell}}, \bibinfo {author}
  {\bibfnamefont {P.~G.}\ \bibnamefont {Radaelli}}, \bibinfo {author}
  {\bibfnamefont {Y.}~\bibnamefont {Singh}}, \bibinfo {author} {\bibfnamefont
  {P.}~\bibnamefont {Gegenwart}}, \bibinfo {author} {\bibfnamefont {K.~R.}\
  \bibnamefont {Choi}}, \bibinfo {author} {\bibfnamefont {S.-W.}\ \bibnamefont
  {Cheong}}, \bibinfo {author} {\bibfnamefont {P.~J.}\ \bibnamefont {Baker}},
  \bibinfo {author} {\bibfnamefont {C.}~\bibnamefont {Stock}}, \ and\ \bibinfo
  {author} {\bibfnamefont {J.}~\bibnamefont {Taylor}},\ }\href {\doibase
  10.1103/PhysRevLett.108.127204} {\bibfield  {journal} {\bibinfo  {journal}
  {Phys. Rev. Lett.}\ }\textbf {\bibinfo {volume} {108}},\ \bibinfo {pages}
  {127204} (\bibinfo {year} {2012})}\BibitemShut {NoStop}%
\bibitem [{\citenamefont {Schirmer}\ \emph {et~al.}(1984)\citenamefont
  {Schirmer}, \citenamefont {Forster}, \citenamefont {Hesse}, \citenamefont
  {Wohlecke},\ and\ \citenamefont {Kapphan}}]{ir_soc_exp}%
  \BibitemOpen
  \bibfield  {author} {\bibinfo {author} {\bibfnamefont {O.~F.}\ \bibnamefont
  {Schirmer}}, \bibinfo {author} {\bibfnamefont {A.}~\bibnamefont {Forster}},
  \bibinfo {author} {\bibfnamefont {H.}~\bibnamefont {Hesse}}, \bibinfo
  {author} {\bibfnamefont {M.}~\bibnamefont {Wohlecke}}, \ and\ \bibinfo
  {author} {\bibfnamefont {S.}~\bibnamefont {Kapphan}},\ }\href {\doibase
  10.1088/0022-3719/17/7/024} {\bibfield  {journal} {\bibinfo  {journal}
  {Journal of Physics C: Solid State Physics}\ }\textbf {\bibinfo {volume}
  {17}},\ \bibinfo {pages} {1321} (\bibinfo {year} {1984})}\BibitemShut
  {NoStop}%
\bibitem [{\citenamefont {Lv}\ \emph {et~al.}(2015)\citenamefont {Lv},
  \citenamefont {Weng}, \citenamefont {Fu}, \citenamefont {Wang}, \citenamefont
  {Miao}, \citenamefont {Ma}, \citenamefont {Richard}, \citenamefont {Huang},
  \citenamefont {Zhao}, \citenamefont {Chen}, \citenamefont {Fang},
  \citenamefont {Dai}, \citenamefont {Qian},\ and\ \citenamefont
  {Ding}}]{lvExperimental2015}%
  \BibitemOpen
  \bibfield  {author} {\bibinfo {author} {\bibfnamefont {B.~Q.}\ \bibnamefont
  {Lv}}, \bibinfo {author} {\bibfnamefont {H.~M.}\ \bibnamefont {Weng}},
  \bibinfo {author} {\bibfnamefont {B.~B.}\ \bibnamefont {Fu}}, \bibinfo
  {author} {\bibfnamefont {X.~P.}\ \bibnamefont {Wang}}, \bibinfo {author}
  {\bibfnamefont {H.}~\bibnamefont {Miao}}, \bibinfo {author} {\bibfnamefont
  {J.}~\bibnamefont {Ma}}, \bibinfo {author} {\bibfnamefont {P.}~\bibnamefont
  {Richard}}, \bibinfo {author} {\bibfnamefont {X.~C.}\ \bibnamefont {Huang}},
  \bibinfo {author} {\bibfnamefont {L.~X.}\ \bibnamefont {Zhao}}, \bibinfo
  {author} {\bibfnamefont {G.~F.}\ \bibnamefont {Chen}}, \bibinfo {author}
  {\bibfnamefont {Z.}~\bibnamefont {Fang}}, \bibinfo {author} {\bibfnamefont
  {X.}~\bibnamefont {Dai}}, \bibinfo {author} {\bibfnamefont {T.}~\bibnamefont
  {Qian}}, \ and\ \bibinfo {author} {\bibfnamefont {H.}~\bibnamefont {Ding}},\
  }\href {https://link.aps.org/doi/10.1103/PhysRevX.5.031013} {\bibfield
  {journal} {\bibinfo  {journal} {Phys. Rev. X}\ }\textbf {\bibinfo {volume}
  {5}} (\bibinfo {year} {2015})}\BibitemShut {NoStop}%
\bibitem [{\citenamefont {Zheng}\ \emph {et~al.}(2022)\citenamefont {Zheng},
  \citenamefont {Gu}, \citenamefont {Liu}, \citenamefont {Tong}, \citenamefont
  {Zhang}, \citenamefont {Zhang}, \citenamefont {Jia}, \citenamefont {Feng},\
  and\ \citenamefont {Du}}]{zhengObservation2022}%
  \BibitemOpen
  \bibfield  {author} {\bibinfo {author} {\bibfnamefont {X.}~\bibnamefont
  {Zheng}}, \bibinfo {author} {\bibfnamefont {Q.}~\bibnamefont {Gu}}, \bibinfo
  {author} {\bibfnamefont {Y.}~\bibnamefont {Liu}}, \bibinfo {author}
  {\bibfnamefont {B.}~\bibnamefont {Tong}}, \bibinfo {author} {\bibfnamefont
  {J.-F.}\ \bibnamefont {Zhang}}, \bibinfo {author} {\bibfnamefont
  {C.}~\bibnamefont {Zhang}}, \bibinfo {author} {\bibfnamefont
  {S.}~\bibnamefont {Jia}}, \bibinfo {author} {\bibfnamefont {J.}~\bibnamefont
  {Feng}}, \ and\ \bibinfo {author} {\bibfnamefont {R.-R.}\ \bibnamefont
  {Du}},\ }\href {\doibase 10.1093/nsr/nwab191} {\bibfield  {journal} {\bibinfo
   {journal} {National Science Review}\ }\textbf {\bibinfo {volume} {9}},\
  \bibinfo {pages} {nwab191} (\bibinfo {year} {2022})}\BibitemShut {NoStop}%
\bibitem [{\citenamefont {Inoue}\ \emph {et~al.}(2016)\citenamefont {Inoue},
  \citenamefont {Gyenis}, \citenamefont {Wang}, \citenamefont {Li},
  \citenamefont {Oh}, \citenamefont {Jiang}, \citenamefont {Ni}, \citenamefont
  {Bernevig},\ and\ \citenamefont {Yazdani}}]{inoueQuasiparticle2016}%
  \BibitemOpen
  \bibfield  {author} {\bibinfo {author} {\bibfnamefont {H.}~\bibnamefont
  {Inoue}}, \bibinfo {author} {\bibfnamefont {A.}~\bibnamefont {Gyenis}},
  \bibinfo {author} {\bibfnamefont {Z.}~\bibnamefont {Wang}}, \bibinfo {author}
  {\bibfnamefont {J.}~\bibnamefont {Li}}, \bibinfo {author} {\bibfnamefont
  {S.~W.}\ \bibnamefont {Oh}}, \bibinfo {author} {\bibfnamefont
  {S.}~\bibnamefont {Jiang}}, \bibinfo {author} {\bibfnamefont
  {N.}~\bibnamefont {Ni}}, \bibinfo {author} {\bibfnamefont {B.~A.}\
  \bibnamefont {Bernevig}}, \ and\ \bibinfo {author} {\bibfnamefont
  {A.}~\bibnamefont {Yazdani}},\ }\href {\doibase 10.1126/science.aad8766}
  {\bibfield  {journal} {\bibinfo  {journal} {Science}\ }\textbf {\bibinfo
  {volume} {351}},\ \bibinfo {pages} {1184} (\bibinfo {year}
  {2016})}\BibitemShut {NoStop}%
\bibitem [{\citenamefont {Huang}\ \emph {et~al.}(2015)\citenamefont {Huang},
  \citenamefont {Zhao}, \citenamefont {Long}, \citenamefont {Wang},
  \citenamefont {Chen}, \citenamefont {Yang}, \citenamefont {Liang},
  \citenamefont {Xue}, \citenamefont {Weng}, \citenamefont {Fang},
  \citenamefont {Dai},\ and\ \citenamefont {Chen}}]{huangchiral2015}%
  \BibitemOpen
  \bibfield  {author} {\bibinfo {author} {\bibfnamefont {X.}~\bibnamefont
  {Huang}}, \bibinfo {author} {\bibfnamefont {L.}~\bibnamefont {Zhao}},
  \bibinfo {author} {\bibfnamefont {Y.}~\bibnamefont {Long}}, \bibinfo {author}
  {\bibfnamefont {P.}~\bibnamefont {Wang}}, \bibinfo {author} {\bibfnamefont
  {D.}~\bibnamefont {Chen}}, \bibinfo {author} {\bibfnamefont {Z.}~\bibnamefont
  {Yang}}, \bibinfo {author} {\bibfnamefont {H.}~\bibnamefont {Liang}},
  \bibinfo {author} {\bibfnamefont {M.}~\bibnamefont {Xue}}, \bibinfo {author}
  {\bibfnamefont {H.}~\bibnamefont {Weng}}, \bibinfo {author} {\bibfnamefont
  {Z.}~\bibnamefont {Fang}}, \bibinfo {author} {\bibfnamefont {X.}~\bibnamefont
  {Dai}}, \ and\ \bibinfo {author} {\bibfnamefont {G.}~\bibnamefont {Chen}},\
  }\href {\doibase 10.1103/PhysRevX.5.031023} {\bibfield  {journal} {\bibinfo
  {journal} {Phys. Rev. X}\ }\textbf {\bibinfo {volume} {5}},\ \bibinfo {pages}
  {031023} (\bibinfo {year} {2015})}\BibitemShut {NoStop}%
\bibitem [{\citenamefont {Osterhoudt}\ \emph {et~al.}(2019)\citenamefont
  {Osterhoudt}, \citenamefont {Diebel}, \citenamefont {Gray}, \citenamefont
  {Yang}, \citenamefont {Stanco}, \citenamefont {Huang}, \citenamefont {Shen},
  \citenamefont {Ni}, \citenamefont {Moll}, \citenamefont {Ran},\ and\
  \citenamefont {Burch}}]{osterhoudt2019}%
  \BibitemOpen
  \bibfield  {author} {\bibinfo {author} {\bibfnamefont {G.~B.}\ \bibnamefont
  {Osterhoudt}}, \bibinfo {author} {\bibfnamefont {L.~K.}\ \bibnamefont
  {Diebel}}, \bibinfo {author} {\bibfnamefont {M.~J.}\ \bibnamefont {Gray}},
  \bibinfo {author} {\bibfnamefont {X.}~\bibnamefont {Yang}}, \bibinfo {author}
  {\bibfnamefont {J.}~\bibnamefont {Stanco}}, \bibinfo {author} {\bibfnamefont
  {X.}~\bibnamefont {Huang}}, \bibinfo {author} {\bibfnamefont
  {B.}~\bibnamefont {Shen}}, \bibinfo {author} {\bibfnamefont {N.}~\bibnamefont
  {Ni}}, \bibinfo {author} {\bibfnamefont {P.~J.~W.}\ \bibnamefont {Moll}},
  \bibinfo {author} {\bibfnamefont {Y.}~\bibnamefont {Ran}}, \ and\ \bibinfo
  {author} {\bibfnamefont {K.~S.}\ \bibnamefont {Burch}},\ }\href {\doibase
  10.1038/s41563-019-0297-4} {\bibfield  {journal} {\bibinfo  {journal} {Nature
  Materials}\ }\textbf {\bibinfo {volume} {18}},\ \bibinfo {pages} {471}
  (\bibinfo {year} {2019})}\BibitemShut {NoStop}%
\bibitem [{\citenamefont {Shanavas}\ \emph {et~al.}(2014)\citenamefont
  {Shanavas}, \citenamefont {Popovi\ifmmode~\acute{c}\else \'{c}\fi{}},\ and\
  \citenamefont {Satpathy}}]{Shanavas2014}%
  \BibitemOpen
  \bibfield  {author} {\bibinfo {author} {\bibfnamefont {K.~V.}\ \bibnamefont
  {Shanavas}}, \bibinfo {author} {\bibfnamefont {Z.~S.}\ \bibnamefont
  {Popovi\ifmmode~\acute{c}\else \'{c}\fi{}}}, \ and\ \bibinfo {author}
  {\bibfnamefont {S.}~\bibnamefont {Satpathy}},\ }\href {\doibase
  10.1103/PhysRevB.90.165108} {\bibfield  {journal} {\bibinfo  {journal} {Phys.
  Rev. B}\ }\textbf {\bibinfo {volume} {90}},\ \bibinfo {pages} {165108}
  (\bibinfo {year} {2014})}\BibitemShut {NoStop}%
\bibitem [{\citenamefont {Sancho}\ \emph {et~al.}(1985)\citenamefont {Sancho},
  \citenamefont {Sancho}, \citenamefont {Sancho},\ and\ \citenamefont
  {Rubio}}]{sanchoHighly1985}%
  \BibitemOpen
  \bibfield  {author} {\bibinfo {author} {\bibfnamefont {M.~P.~L.}\
  \bibnamefont {Sancho}}, \bibinfo {author} {\bibfnamefont {J.~M.~L.}\
  \bibnamefont {Sancho}}, \bibinfo {author} {\bibfnamefont {J.~M.~L.}\
  \bibnamefont {Sancho}}, \ and\ \bibinfo {author} {\bibfnamefont
  {J.}~\bibnamefont {Rubio}},\ }\href {\doibase 10.1088/0305-4608/15/4/009}
  {\bibfield  {journal} {\bibinfo  {journal} {Journal of Physics F: Metal
  Physics}\ }\textbf {\bibinfo {volume} {15}},\ \bibinfo {pages} {851}
  (\bibinfo {year} {1985})}\BibitemShut {NoStop}%
\end{thebibliography}%

%\clearpage
%
%\appendix
%
%\renewcommand{\thefigure}{S-\arabic{figure}}
%\setcounter{figure}{0}
%
%\renewcommand{\theequation}{A-\arabic{equation}}
%\setcounter{equation}{0}
%
%\section*{Supplementary materials}

%\section*{Appendix A.}
%\begin{figure}[ht]
%	\includegraphics[width=70 mm]{./FigS1.png}
%	\caption{XXX}
%	\label{fig:s1}
%\end{figure}

%\section*{Appendix B.}

\end{document}